\documentclass[10pt, conference, letterpaper]{ IEEEtran}
\IEEEoverridecommandlockouts
\usepackage{cite}
\usepackage{amsmath,amssymb,amsfonts}
\usepackage{algorithmic}
\usepackage{booktabs}
\usepackage{tabularx}
\usepackage{booktabs}
\usepackage{graphicx}
\usepackage{subcaption}
\usepackage{textcomp}
\usepackage{xcolor,enumitem}
\usepackage{tikz}
\usetikzlibrary{arrows.meta,positioning,fit,calc}
\usepackage{amsmath}
\usepackage{xcolor}

\usetikzlibrary{
    arrows.meta,
    calc,
    positioning
}

\definecolor{clientblue}{RGB}{32,91,180}
\definecolor{serverred}{RGB}{205,52,52}

\usepackage{xspace}
\newcommand{\paragraphB}[1]{\vskip 4pt \noindent \textit{\textbf{#1}.}\xspace}
\usepackage[most]{tcolorbox}
\tcbset{
    colback=gray!10,       % background color
    colframe=black!70,     % border color
    boxrule=0.4pt,         % border thickness
    arc=2pt,               % rounded corners
    left=6pt, right=6pt, top=4pt, bottom=4pt,
    fonttitle=\bfseries,
    coltitle=black
}

\usepackage{url}
\usepackage[color=green]{todonotes}

\begin{document}

\title{A Large-Scale Benchmark and Risk Assessment of Traffic Analysis Attacks on Cloud LLM Services}

\author{
\IEEEauthorblockN{
Shahrooz Pouryousef\IEEEauthorrefmark{1},
Jesus Lopez\IEEEauthorrefmark{2},
Saeefa Rubaiyet Nowmi\IEEEauthorrefmark{3},
Md Mahmuduzzaman Kamol\IEEEauthorrefmark{3}
}
\IEEEauthorblockN{
Moinul Hossain\IEEEauthorrefmark{4},
Muoi Tran\IEEEauthorrefmark{1},
Mohammad Saidur Rahman\IEEEauthorrefmark{2}
}
\IEEEauthorblockA{
\IEEEauthorrefmark{1}Chalmers University of Technology, Sweden \quad
\IEEEauthorrefmark{2}University of Texas at El Paso, USA
}
\IEEEauthorblockA{
\IEEEauthorrefmark{3}Old Dominion University, USA \quad
\IEEEauthorrefmark{4}George Mason University, USA
}
}

% Shahrooz Pouryousef, Jesus Lopez, Saeefa Rubaiyat Nowmi, Md Mahmuduzzaman Kamol, Moinul Hossain, Muoi Tran, Mohammad Saidur Rahman

\maketitle

% \begin{abstract}

\begin{abstract}
Cloud-based Large language model (LLM) services create a network-level traffic side channel that can expose model, prompt, and task behavior despite encryption. From packet sizes, directions, timing, and burst structure alone, a passive local observer can infer the serving model, the user's prompt category, and the task executed by a collaborative multi-agent system. Yet current evidence is fragmented across separate datasets and settings, limiting reproducibility and comparison. We present, to our knowledge, the first unified measurement study and public benchmark of encrypted LLM traffic across both user--LLM and multi-agent executions. The large-scale benchmark contains 60,000 user--LLM interactions across 10 models and 6 prompt categories, plus 2,838 multi-agent executions covering 10 task categories and two coordination topologies. Using only encrypted packet metadata, we assess the risk of traffic analysis attack by characterizing traffic signatures, identifying the features most associated with leakage, and testing robustness under prompt reformulation, decoding-temperature changes, larger candidate model sets, and partial traffic observation. Model fingerprinting achieves 97.7\% balanced accuracy, prompt-category fingerprinting reaches 76.7\% mean accuracy, and multi-agent task fingerprinting achieves up to 90.7\% accuracy. Prompt reformulation weakens but does not remove model-specific leakage, and task fingerprints remain detectable even from a single agent's traffic. 
%We release PCAP traces, anonymized labels, and evaluation pipelines.
\end{abstract}

\vspace{-0.05in}

\noindent \textbf{Availability.} Code and data are available at \url{https://github.com/IQSeC-Lab/AI-Traffic-Analysis}.

\vspace{-0.1in}
\section{Introduction}%\vspace{-0.05in}
\label{sec:introduction}

Modern large language models (LLMs) are deployed as networked applications. These applications include direct user--LLM interactions where users submit prompts to remote models and collaborative multi-agent workflows, in which multiple agents repeatedly invoke remote models, exchange intermediate results, and coordinate to complete tasks~\cite{wu2024autogen,zhu2025multiagentbench}. We study both settings, focusing on the encrypted network traffic generated during remote LLM inference. Prior research on encrypted traffic analysis has shown that, although encryption prevents direct inspection of communication content, a passive on-path observer can exploit side-channel metadata, e.g., packet sizes, directions, timing, and burst structure, to infer application and user activities~\cite{sirinam2018deep,rahman2020tik,alhazbi2025rhythm}.

Networked LLM services inherit many security risks of distributed systems, including threats to confidentiality, integrity, and availability, but in this paper we focus specifically on privacy leakage to a passive on-path adversary. Such an adversary, such as a local eavesdropper, an ISP, or a network middlebox, records packet metadata without modifying or decrypting the communication. It has been studied extensively in website fingerprinting and traffic analysis, where packet sequences and timing can reveal fine-grained user activity despite transport-layer encryption{~\cite{cai2012touching,dyer2012peek,wang2014effective,hayes2016k,sirinam2018deep,rahman2020tik}.

\textbf{Motivation.} Side-channel fingerprinting attacks have targeted LLM-based services too. Prior work has reconstructed portions of LLM responses and inferred their topics from token-length patterns exposed in encrypted traffic~\cite{jeong2025network,weiss2024prompt} and exploited data-dependent inference timing to infer properties of user inputs~\cite{carlini2024remote}. For model identification, LLMmap actively probes applications and analyzes their outputs~\cite{pasquini2025llmmap}, whereas \emph{LLMs Have Rhythm} passively fingerprints serving models using inter-token timing and encrypted network traffic~\cite{alhazbi2025rhythm}. Other studies infer prompt topics from packet sizes and timing in encrypted streaming responses~\cite{mcdonald2025whisper}, and use encrypted user--agent traffic to identify agent behaviors and agents and to profile user attributes from their inferred agent usage~\cite{zhang2025exposing}. However, these efforts use separate datasets, collection setups, labeling schemes, and adversary assumptions, typically focusing on a narrow set of inference targets or interaction settings. This fragmentation makes it impossible to compare attacks and defenses under a common passive-observer model or to characterize privacy leakage across both direct user--LLM interactions and collaborative multi-agent workflows.

Shared benchmarks helped address this problem in related fields. Tor website-fingerprinting datasets allow researchers to compare attacks and defenses under common conditions, while GTT23 shows why realistic traces are important for measuring practical leakage~\cite{jansen2024measurement}. Similarly, HELM, MMLU, RouterBench, and LLMRouterBench provide common tasks and evaluation settings for comparing LLMs and routing methods~\cite{liang2022holistic,hendrycks2020measuring,hu2024routerbench,li2026llmrouterbench}. However, LLM traffic analysis lacks a similar foundation. To our knowledge, no public benchmark jointly covers direct user--LLM interactions and collaborative multi-agent executions while supporting model, prompt, and task inference under a common passive-observer model. This paper addresses that gap by constructing such a benchmark and assessing the risk of privacy leakage. 

\textbf{Challenges}. Designing such a benchmark is challenging. It must capture real encrypted traffic from both direct user--LLM interactions and collaborative multi-agent workflows, span diverse models, prompts, and tasks, and provide reliable labels at multiple levels of behavior, while restricting the observer to packet-level metadata available to an on-path adversary. For risk assessment, it must support controlled evaluation under changes to prompts, decoding settings, model sets, and traffic visibility to determine if observed leakage reflects persistent execution patterns or artifacts of a particular configuration.

\textbf{Contributions.} To address these challenges, we construct a unified benchmark of encrypted traffic from networked LLM services. It covers $60,000$ direct user--LLM interactions across $10$ models and $6$ prompt categories, together with $2,838$ successful multi-agent executions from MultiAgentBench~\cite{zhu2025multiagentbench} across $10$ task categories and graph and star coordination topologies. All traces are collected and labeled through the same pipeline under a common passive-observer model, providing a consistent basis for studying leakage. % across models, prompts, tasks, and coordination topologies.

We use the benchmark to evaluate both established and less explored inference tasks. For established problems, the benchmark supports serving-model and prompt-category fingerprinting under a common evaluation setting. We also study multi-agent task fingerprinting, where an observer infers the collaborative task. These experiments show that the benchmark supports model-, prompt-, and task-level analysis and enables consistent evaluation across interaction settings.

Finally, we assess whether the observed leakage persists under changes in execution and observation. We vary prompt wording, decoding temperature, candidate model sets, and the amount of visible agent--LLM traffic. Model fingerprinting achieves $97.7\%$ balanced accuracy, prompt-category fingerprinting reaches $76.7\%$ mean accuracy, and multi-agent task fingerprinting achieves up to $90.7\%$ accuracy. Prompt reformulation weakens but does not eliminate model-specific leakage, while collaborative tasks remain detectable even when the observer sees traffic from only one participating agent. These results show that the observed leakage persists beyond a single configuration. In summary, the key contributions are:

\begin{itemize}[leftmargin=*]
    \item \textbf{Unified encrypted-traffic benchmark.} We construct a benchmark of $60,000$ user--LLM interactions and $2,838$ multi-agent executions spanning diverse models, prompts, tasks, and coordination topologies.
    
    \item \textbf{Risk assessment across inference tasks.} We use the benchmark for serving-model and prompt-category fingerprinting and for the less-explored task of inferring multi-agent tasks. We also identify the features associated with each target.
    
    \item \textbf{Robustness under realistic variation.} We test prompt reformulation, temperature changes, larger model sets, and partial traffic visibility, showing when attacks weaken and when leakage persists.
\end{itemize}

\section{Related Work}%\vspace{-0.02in}

\noindent\textbf{Encrypted-traffic fingerprinting.}
Encrypted connections expose metadata, e.g., packet sizes, directions, timing, and burst structure. Website fingerprinting (WF) showed that these signals can identify user activity even over Tor~\cite{hintz2002fingerprinting,cai2012touching,wang2014effective}. Early attacks relied on hand-crafted features, whereas deep-learning methods learned representations directly from packet sequences and timing~\cite{hayes2016k,sirinam2018deep,Sirinam2019,rahman2020tik}. Later studies examined actual traffic, changing network conditions, and adaptive defenses, showing that lab performance may not transfer directly to realistic settings~\cite{cherubin2022online,bahramali2023realistic,mathews2023sok}. This literature motivates our use of encrypted packet metadata, but our objective is to infer properties of LLM computation rather than visited websites.

\noindent\textbf{Traffic analysis of LLM services.}
Autoregressive generation and response streaming create observable patterns
that reveal conversation content and serving-model properties.
Weiss~\cite{weiss2024prompt} reconstructs generated text and infers conversation
topics from encrypted response sizes. Whisper Leak infers target topics from
packet sizes and timing, while NetEcho reconstructs prompts and responses using
offline probing and trace--text matching~\cite{mcdonald2025whisper,zhang2025netecho}.
Alhazbi~\cite{alhazbi2025llms} shows that inter-token timing produces
model-specific signatures, LLMmap identifies serving models using an active
adversary with plaintext responses~\cite{pasquini2025llmmap}, and Carlini and
Nasr expose remotely observable timing leakage caused by data-dependent
inference optimizations~\cite{carlini2024remote}. These works demonstrate
leakage of topics, content, model identity, and serving behavior, but typically
study a single target. Our benchmark instead supports serving-model and
multiclass prompt-category fingerprinting under a common passive-observer model
and evaluates robustness to prompt reformulation, decoding temperature, larger
candidate sets, and partial traffic visibility.

\noindent\textbf{Traffic analysis of LLM agents.}
Agentic systems generate rich traffic through planning, tool use, retrieval, and repeated model calls. AgentPrint analyzes encrypted user--agent traffic to identify agent activities and applications and infer user attributes from longitudinal usage patterns~\cite{zhang2025exposing}. PersonaFingerprint shows that encrypted traffic produced by LLM-driven browsing can expose high-level user personas~\cite{song2026personafingerprint}. These studies examine individual agents or their external browsing behavior. In contrast, we study encrypted traffic jointly generated by collaborative multi-agent executions and assess if a passive observer can infer the task being performed. We evaluate graph and star coordination topologies and partial visibility, where only a subset of agent--LLM flows is observed. 

\section{Proposed Threat Model}%\vspace{-0.01in}
\label{sec:threat}

\begin{figure*}[t]
    \centering
    \includegraphics[width=0.95\textwidth]{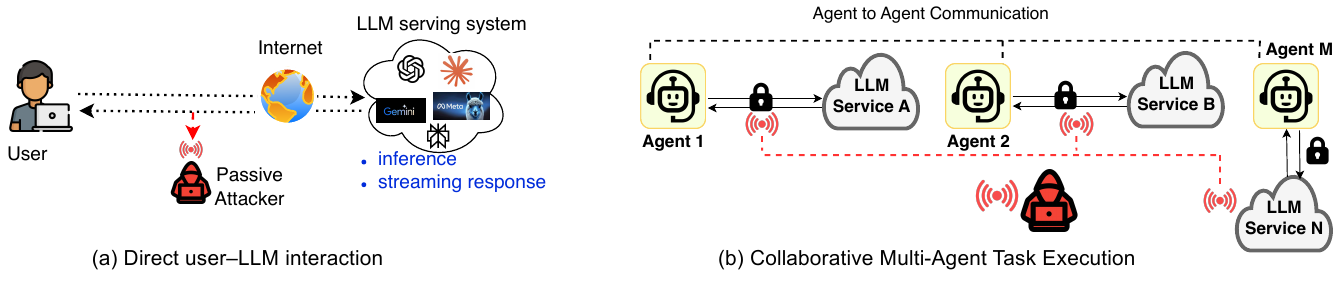}\vspace{-0.05in}
    \caption{\small Figure~\ref{fig:threat-model}(a) illustrates the direct user--LLM setting. Figure~\ref{fig:threat-model}(b) illustrates the collaborative multi-agent setting, where the adversary observes encrypted agent--LLM traffic but not agent-to-agent communication.}
    \label{fig:threat-model}%\vspace{-0.35in}
\end{figure*}

% \begin{figure*}[t]
%     \centering

%     \begin{subfigure}[t]{0.47\textwidth}
%         \centering
%         \includegraphics[width=\textwidth]{img/LLM-a.png}\vspace{-0.1in}
%         \caption{\small Direct user--LLM interaction.}
%         \label{fig:threat-direct}
%     \end{subfigure}
%     \hfill
%     \begin{subfigure}[t]{0.47\textwidth}
%         \centering
%         \includegraphics[width=\textwidth]{img/LLM-b.png}\vspace{-0.1in}
%         \caption{\small Collaborative multi-agent execution.}
%         \label{fig:threat-multi}
%     \end{subfigure}\vspace{-0.05in}

%     \caption{\small\textbf{Threat Model.} Figure~\ref{fig:threat-model}(a) illustrates the direct user--LLM setting. Figure~\ref{fig:threat-model}(b) illustrates the collaborative multi-agent setting, where the adversary observes encrypted agent--LLM traffic but not agent-to-agent communication.
%     }
%     \label{fig:threat-model}\vspace{-0.25in}
% \end{figure*}

\subsection{System and Observation Model}%\vspace{-0.01in}

Figure~\ref{fig:threat-model} summarizes the two considered settings. In the direct setting, a user communicates with a remote LLM service over TLS. A passive on-path adversary records the encrypted traffic exchanged during each interaction; it cannot decrypt or modify the traffic, and neither the endpoint is compromised.

In the collaborative setting, multiple LLM-based agents\footnote{An LLM agent is a software system that combines one or more LLMs with tools, memory, and workflow logic to autonomously plan and execute tasks. Multi-agent systems comprise multiple collaborating LLM agents.} solve a shared task using MARBLE~\cite{zhu2025multiagentbench} under its star and graph-mesh coordination protocols. The topology of the coordination can be star, i.e., a central planner assigns subtasks and combines agent outputs, or mesh, i.e., agents communicate directly. Agents may communicate with the same or different remote LLM services. In both settings, the adversary observes only packet timestamps, sizes, and directions. Packet payloads, prompts, responses, IP addresses, and endpoint identities remain hidden. In the collaborative setting, agent-to-agent communication is also unobservable. An observed traffic trace is represented as
$F=\{(t_i,s_i,d_i)\}_{i=1}^{n}$,
where $t_i$, $s_i$, and $d_i$ denote the relative timestamp, size, and direction of packet $i$, respectively. We assume the adversary can correctly segment each interaction, direct or collaborative.\vspace{-0.02in}

\subsection{Adversary Knowledge, Objectives, and Scope}\vspace{-0.01in}

We consider a supervised closed-world setting. The adversary has access to labeled reference traces for the candidate labels, trains fingerprinting models offline, and predicts the label of an observed trace. For direct user--LLM interactions, the candidate labels correspond to serving models or prompt categories. The adversary seeks to infer
$f_{\mathrm{model}}:F\rightarrow\mathcal{M},
\qquad
f_{\mathrm{category}}:F\rightarrow\mathcal{P}$,
where $\mathcal{M}$ and $\mathcal{P}$ denote the candidate model and prompt-category sets, respectively. For collaborative executions, the candidate labels correspond to collaborative task categories. The adversary seeks to infer
$f_{\mathrm{task}}:F_{\mathrm{multi}}\rightarrow\mathcal{T}$,
where $F_{\mathrm{multi}}$ denotes the aggregated encrypted traffic observed from the monitored agents during a single execution, and $\mathcal{T}$ is the set of candidate task categories. Coordination topology is not considered an inference target.\vspace{-0.05in}

% \subsection{Observation Scope}

%% MSR: need to validate with experimental results
%For collaborative executions, we consider two observation settings. In \emph{global observation}, the adversary monitors the encrypted traffic of all participating agents, whereas in \emph{partial observation}, it monitors only a subset. 
%The impact of partial observation and different coordination topologies is evaluated in Section~\ref{sec:results}.

\subsection{Traffic Fingerprinting Workflow}

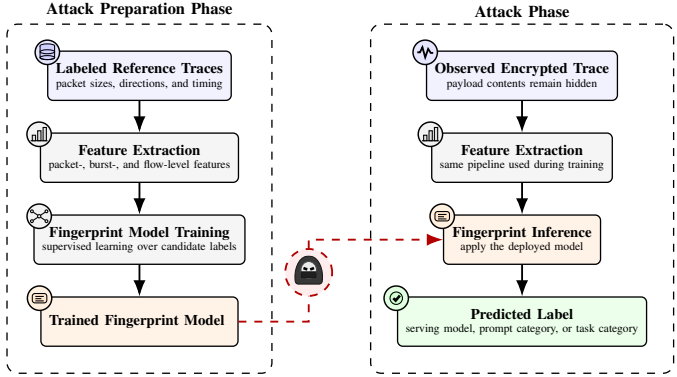
\begin{figure}
    \centering
    \begin{tikzpicture}[
    font=\small,
    >=Latex,
    scale=0.62,
    transform shape,
    node distance=0.68cm and 1.35cm,
    box/.style={
        draw,
        rounded corners=2pt,
        align=center,
        minimum width=3.35cm,
        minimum height=1.05cm,
        inner sep=5pt,
        line width=0.5pt
    },
    data/.style={
        box,
        fill=blue!5
    },
    process/.style={
        box,
        fill=gray!8
    },
    model/.style={
        box,
        fill=orange!10
    },
    output/.style={
        box,
        fill=green!8
    },
    phase/.style={
        draw,
        dashed,
        rounded corners=4pt,
        inner sep=10pt,
        line width=0.5pt
    },
    flow/.style={
        ->,
        line width=0.7pt
    },
    deploy/.style={
        ->,
        dashed,
        line width=0.7pt,
        draw=red!70!black
    },
    attacker/.style={
        circle,
        draw=red!70!black,
        fill=red!7,
        minimum size=10mm,
        inner sep=1pt,
        line width=0.7pt
    }
]

% =========================================================
% Offline attack-preparation column
% =========================================================

\node[data] (reference) {
    \textbf{Labeled Reference Traces}\\[-0.4ex]
    {\scriptsize packet sizes, directions, and timing}
};

\node[process,below=of reference] (features-train) {
    \textbf{Feature Extraction}\\[-0.4ex]
    {\scriptsize packet-, burst-, and flow-level features}
};

\node[process,below=of features-train] (training) {
    \textbf{Fingerprint Model Training}\\[-0.4ex]
    {\scriptsize supervised learning over candidate labels}
};

\node[model,below=of training] (trained) {
    \textbf{Trained Fingerprint Model}
};

\draw[flow] (reference) -- (features-train);
\draw[flow] (features-train) -- (training);
\draw[flow] (training) -- (trained);

\node[
    phase,
    fit=(reference)(features-train)(training)(trained),
    label={[font=\bfseries]above:Attack Preparation Phase}
] (offline) {};

% =========================================================
% Online attack column
% =========================================================

\node[data,right=4.15cm of reference] (observed) {
    \textbf{Observed Encrypted Trace}\\[-0.4ex]
    {\scriptsize payload contents remain hidden}
};

\node[process,below=of observed] (features-test) {
    \textbf{Feature Extraction}\\[-0.4ex]
    {\scriptsize same pipeline used during training}
};

\node[model,below=of features-test] (inference) {
    \textbf{Fingerprint Inference}\\[-0.4ex]
    {\scriptsize apply the deployed model}
};

\node[output,below=of inference] (prediction) {
    \textbf{Predicted Label}\\[-0.4ex]
    {\scriptsize serving model, prompt category, or task category}
};

\draw[flow] (observed) -- (features-test);
\draw[flow] (features-test) -- (inference);
\draw[flow] (inference) -- (prediction);

\node[
    phase,
    fit=(observed)(features-test)(inference)(prediction),
    label={[font=\bfseries]above:Attack Phase}
] (online) {};

% =========================================================
% Deployment connection with attacker icon
% =========================================================

\draw[deploy]
    (trained.east)
    -- ++(1.5,0)
    |- node[pos=0.30,attacker] (attacker-icon) {}
    (inference.west);

% Hooded attacker icon
\begin{scope}[shift={(attacker-icon.center)},scale=0.45]

    % Hood
    \draw[
        fill=black!78,
        draw=black!90,
        rounded corners=1pt
    ]
        (-0.62,-0.55)
        .. controls (-0.72,0.12) and (-0.48,0.70) ..
        (0,0.78)
        .. controls (0.48,0.70) and (0.72,0.12) ..
        (0.62,-0.55)
        -- cycle;

    % Face
    \draw[
        fill=gray!15,
        draw=black!70
    ]
        (0,0.12) ellipse (0.37 and 0.40);

    % Angry eyes
    \draw[line width=1pt]
        (-0.25,0.21) -- (-0.07,0.14);

    \draw[line width=1pt]
        (0.25,0.21) -- (0.07,0.14);

    % Face mask
    \draw[
        fill=black!88,
        draw=black!90
    ]
        (-0.34,0.03) rectangle (0.34,-0.24);

\end{scope}

% =========================================================
% Compact icon badges
% =========================================================

\tikzset{
    badge/.style={
        circle,
        draw,
        fill=white,
        minimum size=5.2mm,
        inner sep=2pt,
        line width=0.55pt
    }
}

% Reference-trace database icon
\node[
    badge,
    fill=blue!10
] (ico-reference) at ($(reference.west)+(0.0007,0.5)$) {};

\begin{scope}[shift={(ico-reference.center)},scale=0.25]
    \draw[fill=blue!18]
        (0,0.55) ellipse (0.62 and 0.17);

    \draw[fill=blue!18]
        (-0.62,0.55)
        -- (-0.62,-0.45)
        arc[
            start angle=180,
            end angle=360,
            x radius=0.62,
            y radius=0.17
        ]
        -- (0.62,0.55);

    \draw
        (0,0.05) ellipse (0.62 and 0.17);

    \draw
        (0,-0.45) ellipse (0.62 and 0.17);
\end{scope}

% Feature-extraction icons
\foreach \n/\iname in {
    features-train/ico-ft,
    features-test/ico-fi
}{
    \node[
        badge,
        fill=gray!10
    ] (\iname) at ($(\n.west)+(-0.02,0.5)$) {};

    \begin{scope}[shift={(\iname.center)},scale=0.25]
        \draw
            (-0.72,-0.55) -- (0.72,-0.55);

        \draw[fill=gray!25]
            (-0.58,-0.55) rectangle (-0.30,-0.08);

        \draw[fill=gray!25]
            (-0.12,-0.55) rectangle (0.16,0.30);

        \draw[fill=gray!25]
            (0.34,-0.55) rectangle (0.62,0.58);
    \end{scope}
}

% Model-training icon
\node[
    badge,
    fill=gray!10
] (ico-training) at ($(training.west)+(0.12,0.5)$) {};

\begin{scope}[shift={(ico-training.center)},scale=0.25]
    \draw (-0.62,0.42) circle (0.12);
    \draw (-0.62,-0.42) circle (0.12);
    \draw (0,0) circle (0.14);
    \draw (0.62,0.42) circle (0.12);
    \draw (0.62,-0.42) circle (0.12);

    \draw (-0.50,0.34) -- (-0.12,0.09);
    \draw (-0.50,-0.34) -- (-0.12,-0.09);
    \draw (0.12,0.09) -- (0.50,0.34);
    \draw (0.12,-0.09) -- (0.50,-0.34);
\end{scope}

% Model and inference icons
\foreach \n/\iname in {
    trained/ico-trained,
    inference/ico-inference
}{
    \node[
        badge,
        fill=orange!12
    ] (\iname) at ($(\n.west)+(-0.02,0.5)$) {};

    \begin{scope}[shift={(\iname.center)},scale=0.25]
        \draw[
            rounded corners=1pt,
            fill=orange!20
        ]
            (-0.62,-0.40) rectangle (0.62,0.40);

        \draw
            (-0.36,0.15) -- (0.36,0.15);

        \draw
            (-0.36,-0.04) -- (0.15,-0.04);

        \draw
            (-0.36,-0.23) -- (0.36,-0.23);
    \end{scope}
}

% Observed encrypted-trace icon
\node[
    badge,
    fill=blue!10
] (ico-observed) at ($(observed.west)+(-0.02,0.5)$) {};

\begin{scope}[shift={(ico-observed.center)},scale=0.25]
    \draw[line width=0.7pt]
        (-0.68,0)
        -- (-0.45,0)
        -- (-0.24,0.43)
        -- (0,-0.43)
        -- (0.25,0.29)
        -- (0.47,-0.05)
        -- (0.68,-0.05);
\end{scope}

% Prediction icon
\node[
    badge,
    fill=green!12
] (ico-prediction) at ($(prediction.west)+(-0.02,0.5)$) {};

\begin{scope}[shift={(ico-prediction.center)},scale=0.25]
    \draw[fill=green!18]
        (0,0) circle (0.53);

    \draw[line width=1pt]
        (-0.25,0)
        -- (-0.05,-0.20)
        -- (0.31,0.23);
\end{scope}

\end{tikzpicture}%\vspace{-0.2in}
    \caption{\small\textbf{Workflow  of the traffic-analysis attack.} The adversary trains a fingerprint model via labeled encrypted traffic traces and applies it to observed encrypted traffic to infer the serving model, prompt category, or collaborative task.}\vspace{-0.1in}
    \label{fig:attack-overview}
\end{figure}

Figure~\ref{fig:attack-overview} illustrates the considered traffic fingerprinting workflow. Under the closed-world assumption, the adversary first collects labeled reference traces for the candidate classes and trains a supervised classifier offline. Each traffic trace is segmented into an individual user--LLM interaction or collaborative multi-agent execution, from which packet-level and flow-level features are extracted to construct a feature vector. During deployment, the adversary observes an encrypted traffic trace from an unknown interaction, applies the same segmentation and feature extraction procedure, and provides the resulting feature vector to the trained classifier. Depending on the inference objective, the classifier predicts the serving model, the prompt category, or the collaborative task category. This workflow is common to all experiments, only the prediction target changes across the different inference tasks.

\section{Proposed Experimental Methodology}%\vspace{-0.025in}
\label{sec:methodology}

This section describes the generation of the user--LLM and multi-agent
datasets, the traffic collection and preprocessing pipeline, and the extracted
traffic features.

\vspace{-0.08in}
\subsection{Dataset Generation}\vspace{-0.02in}
\label{sec:dataset_generation}

\paragraphB{User--LLM Dataset}
The dataset contains 60 prompts evenly distributed across 6 semantic categories: summarization, code generation, mathematical reasoning, logical reasoning, technical explanation, and adversarial requests.
% The user–LLM dataset is generated from $60$ prompts distributed evenly across $6$ semantic categories: summarization, code generation, mathematical reasoning, logic puzzles, technical explanation, and adversarial requests. 
Examples of prompts include ``Implement binary search in Python" for code generation, and ``If you have a drawer full of unmatched socks, how many do you have to pull out to get a matching pair?" for logical reasoning.
% (e.g., "Implement binary search in python" for code generation, or "If you have a drawer full of unmatched socks, how many do you have to pull out to get a match?" for logic puzzles). 
Each prompt is independently submitted to every candidate LLM over 100 fresh sessions. The framework automatically records the model identity, prompt ID, prompt category, and trial ID as ground-truth labels.
% Each prompt is submitted independently to every candidate LLM and repeated $100$ times using a fresh interaction session. The collection framework automatically records the model identity, prompt identifier, prompt category, and repetition number as ground-truth labels.

%\textcolor{red}{\emph{[Describe the candidate LLMs, model versions, inference interface, serving platform, generation parameters, collection period, and total number of executions.]}}

\paragraphB{Multi-Agent Dataset}
We construct the multi-agent dataset using MARBLE~\cite{zhu2025multiagentbench}, where teams of LLM-based agents collaborate on shared tasks. It includes 10 categories with 15 tasks per category. Four categories are native to MARBLE: bargaining, research, coding, and database. We additionally incorporate bug fixing from TeamBench~\cite{kim2026teambench}, software-engineering issue resolution from SWE-bench Lite~\cite{jimenez2024swe}, long-form report writing from DeepResearch Bench~\cite{du2025deepresearch}, clinical reasoning from MedQA~\cite{jin2021disease}, contract analysis from LegalBench~\cite{guha2023legalbench}, and argumentation based on competitive
debate motions~\cite{debatemotions}.
% We build our multi-agent dataset using MARBLE~\cite{zhu2025multiagentbench}, in which teams of LLM-backed
% agents collaborate on shared tasks. The dataset covers 10 categories with $15$ distinct tasks each. 4 categories are native to MARBLE: bargaining, research, coding, and database. We add bug fixing from
% TeamBench~\cite{kim2026teambench}, software-engineering issue resolution from SWE-bench Lite~\cite{jimenez2024swe}, long-form report writing from DeepResearch Bench~\cite{du2025deepresearch}, clinical reasoning from MedQA~\cite{jin2021disease}, and contract analysis from
% LegalBench~\cite{guha2023legalbench}.

Each task is executed under two coordination topologies. In the fully
connected graph, agents communicate directly with all peers; in the star
topology, a central planner coordinates all communication. Teams contain 2--5
role-specialized agents. Execution ends when the planner declares completion
or the configured round limit is reached. All agents use the same locally
hosted Llama~3.2 3B model via Ollama, isolating task and topology effects from
model variation.

% Each task is executed under two agent-coordination topologies. The fully connected graph topology permits direct peer-to-peer communication, whereas the star topology routes coordination through a central agent. Teams contain 2-5 role-specialized agents whose roles are fixed by the task configuration. Under star, the central planner assigns work to active agents; under graph, every agent acts independently in each round and may communicate directly with its peers. Executions terminate when the planner determines that the task is complete or when the configured round limit is reached. All agents use the same locally hosted Llama~3.2 3B model served through Ollama, ensuring that observed differences arise from tasks and coordination rather than from underlying LLMs.

Each task is independently repeated up to 15 times to capture variation from stochastic decoding. The database tasks additionally interact with a containerized PostgreSQL system containing realistic performance anomalies. Overall, we obtain $4,070$ successful and $42$ failed executions, corresponding to a $98.9\%$ success rate. Of the successful executions, $3,970$ contain valid encrypted-traffic captures and are retained.

Task categories lead to different communication structures, varying in coordination rounds, LLM request/response size, and the amount of tool or environment activity between calls. These workflows create distinct packet count, size, burst, and timing patterns without requiring access to encrypted content.

%Task categories naturally induce different communication structures: they vary in the number of coordination rounds, the size of LLM requests and responses, and the amount of tool or environment activity between calls. Consequently, their workflows produce distinguishable packet-count, size, burst, and timing patterns without requiring access to encrypted content.

% Table~\ref{tab:marble-category-counts} summarizes the resulting datasets after
% preprocessing.

% \begin{table}[t]
%     \centering
%     \small
%     \begin{tabular}{lrr}
%         \toprule
%         Category & Graph & Star \\
%         \midrule
%         Bargaining         & 201 & 201 \\
%         Bug fixing         & 198 & 198 \\
%         Coding             & 198 & 198 \\
%         Database           & 198 & 198 \\
%         Debate             & 198 & 198 \\
%         Deep research      & 198 & 198 \\
%         Legal review       & 198 & 198 \\
%         Medical diagnosis  & 198 & 198 \\
%         Research           & 200 & 200 \\
%         SWE-bench          & 198 & 198 \\
%         \bottomrule
%     \end{tabular}
%     \caption{Successful, captured executions per category and coordination
%     topology in the multi-agent dataset.}
%     \label{tab:marble-category-counts}
% \end{table}

% \subsubsection{Data Validation}

% Collected traces are validated before preprocessing. Executions associated with
% failed requests, interrupted workflows, incomplete captures, or corrupted
% recordings are discarded.

\vspace{-0.08in}
\subsection{Traffic Collection and Preprocessing}\vspace{-0.01in}
\label{sec:collection}

Traffic is captured with \texttt{tcpdump} at the earliest observable point on the network path, i.e., the client-facing interface for single-agent interactions and the proxy-facing loopback interface for multi-agent executions. Each capture is synchronized with its corresponding execution via recorded start and end timestamps, allowing a continuously running capture to be sliced into per-execution traces after the fact, rather than requiring the capture tool to be restarted for every interaction.

Each trace is anonymized and preprocessed before feature extraction: packets
are restricted to the execution's time window,
timestamps are normalized relative to the first packet of the trace, and
direction is labeled outgoing/incoming from the fixed client/server role of
the connection endpoints. Byte-identical duplicate packets, arising when
overlapping capture processes observe the same traffic, are removed by
content hashing; genuine TCP retransmissions are rare under our capture
conditions and are left unmodified where they occur. Executions whose
capture failed to record any packets due to transient tooling faults were
discarded and re-executed.

\vspace{-0.08in}
\subsection{Experimental Environment}
\label{sec:Environment}
The experimental pipeline, including local LLM inference, packet capture, PCAP processing, and feature extraction, was executed on a high-performance computing server equipped with two Intel Xeon Gold 6430 processors (64 physical cores, 128 logical cores) and 512~GB of RAM. The server contained two NVIDIA H100 NVL GPUs, each with 95.8 GB of memory. GPU-accelerated workloads were executed using CUDA 12.8 and NVIDIA driver version 595.71.05.
%\vspace{-0.05in}
\section{Risk Assessment and Benchmark Evaluation}%\vspace{-0.05in}
\label{sec:evaluation}
In this section, we empirically evaluate the encrypted traffic fingerprinting
attacks introduced in Section~\ref{sec:threat}. First, we characterize the encrypted traffic generated by different models, prompt categories, and collaborative multi-agent tasks to understand whether they exhibit distinguishable communication patterns. Second, we assess the risk of encrypted traffic fingerprinting for model, prompt-category, and multi-agent task identification. Third, we investigate which observable traffic characteristics are responsible for the information leakage. Finally, we evaluate the practicality of the proposed fingerprinting attack.

%\textcolor{red}{Although the evaluated Hugging Face models and prompt categories are identified by name in the text, the figures use consistent anonymized identifiers for both.}

% Although we identify the evaluated Hugging Face models and prompt categories
% in the text, figures use consistent anonymized identifiers for models and categories.

% \subsection{Experimental Setup}

Unless stated otherwise, all experiments use a Random Forest (RF) classifier.
We report balanced accuracy (BA) as the primary evaluation metric together with
the macro-averaged F1 score (Macro-F1). Balanced accuracy is computed as the
average recall across all $C$ classes,%\vspace{-0.05in}

\begin{small}
\begin{equation}
\mathrm{BA}=\frac{1}{C}\sum_{i=1}^{C}
\frac{TP_i}{TP_i+FN_i},
\end{equation}\vspace{-0.075in}
\end{small}

\noindent where $TP_i$ and $FN_i$ denote the true positives and false negatives of class
$i$, respectively. The Macro-F1 is defined as\vspace{-0.075in}

\begin{small}
\begin{equation}
\mathrm{Macro\mbox{-}F1}
=
\frac{1}{C}\sum_{i=1}^{C}
\frac{2P_iR_i}{P_i+R_i},
\end{equation}\vspace{-0.075in}
\end{small}

\noindent where $P_i$ and $R_i$ denote class $i$'s precision and recall.

\begin{figure}[t]
    \centering
    \includegraphics[width=\columnwidth]{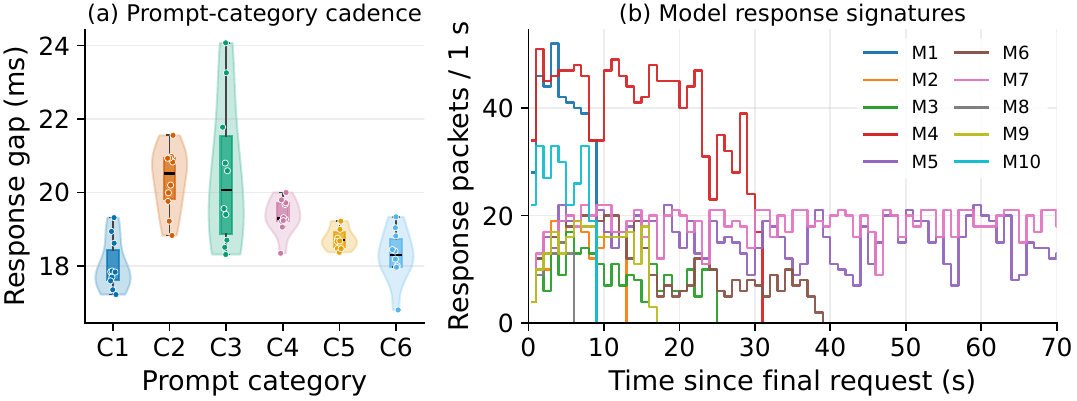}%\vspace{-0.05in}
\caption{\small Encrypted user--LLM traffic properties.
(a) Response-packet inter-arrival timing across prompt categories for a fixed
LLM. Each point represents one prompt; boxes show medians and interquartile
ranges. (b) Temporal response-packet signatures of $10$ LLMs for a fixed prompt.
Each line is the execution closest to the model's median behavior over $100$
trials. Payload-bearing LLM-to-user packets are aggregated into $1$\,s
intervals.}
    \label{fig:user-llm-characteristics}%\vspace{-0.35in}
\end{figure}

\subsection{User--LLM Fingerprinting}

\begin{tcolorbox}[
    colback=blue!5,
    colframe=blue!40!black,
    colbacktitle=blue!15!white,
    coltitle=black,
    fonttitle=\bfseries,
    title filled,
    boxrule=0.5pt,
    arc=2mm,
    left=0.0mm, right=0.1mm, top=0.1mm, bottom=0.01mm,
    % boxrule=0.8pt,
    % arc=2mm,
    % left=1mm, right=4mm, top=2mm, bottom=2mm,
    title={\textit{\small Key Insight: One Traffic Stream, Two Leakage Sources}}
]
\small The serving model shapes how the response is streamed, while the prompt category shapes how much and how long the model generates. One passive capture therefore supports two inference attacks simultaneously, so we assess each risk in isolation by fixing the prompt when fingerprinting models and fixing the model when fingerprinting prompt categories.
% The encrypted response stream simultaneously leaks model identity and prompt category, but through different traffic characteristics. We therefore isolate these two sources of variation by fixing the prompt when fingerprinting models and fixing the model when fingerprinting prompt categories.
\end{tcolorbox}

We evaluate encrypted traffic generated by direct interactions with $10$ LLMs downloaded from Hugging Face: Llama-3.2-3B-SRL, Qwen2.5-7B-Instruct, Qwen3-14B-Base, MegaScience-Qwen3-14B, Phi-4-Mini-Reasoning, AceReason-Nemotron-1.1-7B, PatronusAI-Glider, LLMTwin-Llama-3.1-8B, Typhoon2-8B-Instruct, and XtraGPT-3B. We evaluate $10$ more models to assess scalability. Unless otherwise noted, all experiments use traffic from the original $10$ models.
%We test ten more models for scalability; unless noted, all experiments use traffic from the initial ten models.

\subsubsection{Encrypted Traffic Leaks Distinguishable Signals}

We first examine whether encrypted traffic generated by different LLMs exhibits
distinguishable statistical patterns. For each PCAP trace, our feature-
extraction pipeline computes $176$ content-independent traffic statistics from
packet direction, encrypted payload length, arrival time, rate, bursts, idle
periods, and packet ordering. We exclude IP addresses, endpoint identifiers,
payload contents, and other application metadata. After removing $8$ constant
features, each interaction is represented by a $168$-dimensional feature
vector.  A directional burst is a maximal sequence of consecutive
payload-bearing packets traveling in the same direction,
\[
B=(p_i,\ldots,p_j),
\]
whose duration is
\[
D(B)=t_j-t_i .
\]

Figure~\ref{fig:user-llm-characteristics}(a) shows that prompt category affects encrypted response timing even for a fixed LLM. Categories C2 and C3 generally exhibit longer inter-packet gaps than C1 and C6, while C3 shows greater variation, consistent with math and algorithmic prompts requiring different generation effort. However, the figure cannot separate token-generation time from serving and packetization effects. Figure~\ref{fig:user-llm-characteristics}(b) shows the complementary model-dependent effect for a fixed prompt: models differ in packet rates, response durations, and streaming patterns. These distinct timing and streaming signatures are consistent with prior work~\cite{alhazbi2025rhythm,weiss2024your,zhang2025exposing}.

% Figure~\ref{fig:user-llm-characteristics}(a) shows that prompt category
% affects encrypted response-packet timing even with a fixed LLM. Categories C2
% and C3 generally produce longer response-packet gaps than C1 and C6, while C3
% shows substantially greater variation, consistent with math and algorithmic
% prompts requiring different amounts of reasoning. The figure alone, however,
% does not distinguish token-generation time from serving and packetization
% effects. Figure~\ref{fig:user-llm-characteristics}(b) shows the complementary
% model-dependent effect for a fixed prompt. The evaluated models exhibit
% different packet rates, response durations, and streaming patterns.
% Differences in model architecture, scale, and generation behavior produce
% distinct encrypted timing and streaming patterns, consistent with prior
% work~\cite{alhazbi2025rhythm,weiss2024your,zhang2025agentprint}.

% \noindent\textbf{\textit{Insight}.}
% The same encrypted response stream carries two different signals. For a
% fixed prompt, changing the deployed model alters response duration, packet production, and streaming behavior; for a fixed model, changing the prompt category alters the response-packet gaps. This motivates evaluating the two attacks conditionally: we hold the prompt set fixed when fingerprinting models and train a separate classifier for each model when fingerprinting prompt categories, thereby preventing the two sources of variation from being conflated.

\begin{figure}[t]
    \centering
    \resizebox{\columnwidth}{!}{%
        \begin{tikzpicture}[
    font=\small,
    >=Latex,
    packet/.style={
        draw,
        minimum width=0.17cm,
        minimum height=0.44cm,
        inner sep=0pt,
        line width=0.45pt
    },
    cpacket/.style={
        packet,
        draw=clientblue,
        fill=clientblue!20
    },
    spacket/.style={
        packet,
        draw=serverred,
        fill=serverred!20
    },
    burstbox/.style={
        rounded corners=3pt,
        dashed,
        line width=0.65pt,
        minimum width=1.85cm,
        minimum height=1.85cm,
        text width=1.55cm,
        align=center,
        inner sep=5pt
    },
    cburst/.style={
        burstbox,
        draw=clientblue
    },
    sburst/.style={
        burstbox,
        draw=serverred
    },
    directionlabel/.style={
        font=\bfseries,
        inner sep=1pt
    },
    sliceline/.style={
        black,
        densely dotted,
        line width=1.15pt
    }
]

% =================================================
% DIRECTION LABELS
% =================================================

\node[
    directionlabel,
    anchor=center,
    text=clientblue
] at (3.25,4.30)
    {Client $\rightarrow$ Server};

\node[
    directionlabel,
    anchor=center,
    text=serverred
] at (7.45,4.30)
    {Server $\rightarrow$ Client};

% =================================================
% TIME AXIS
% =================================================

\draw[
    ->,
    line width=0.8pt
] (0.35,3.35) -- (10.65,3.35)
    node[right] {Time};

% =================================================
% PACKET SEQUENCE
% =================================================

% Burst 1: Client to server
\foreach \x in {1.00,1.30,1.60,1.90}
    \node[cpacket,anchor=south] at (\x,3.35) {};

% Burst 2: Server to client
\foreach \x in {2.85,3.15,3.45}
    \node[spacket,anchor=north] at (\x,3.35) {};

% Burst 3: Client to server
\foreach \x in {4.75,5.05,5.35,5.65}
    \node[cpacket,anchor=south] at (\x,3.35) {};

% Burst 4: Server to client
\foreach \x in {6.85,7.15,7.45,7.75,8.05,8.35}
    \node[spacket,anchor=north] at (\x,3.35) {};

% Burst 5: Client to server
\foreach \x in {9.25,9.55,9.85}
    \node[cpacket,anchor=south] at (\x,3.35) {};

% =================================================
% BURST-SLICING BOUNDARIES
% =================================================

\draw[sliceline]
    (2.50,4.02) -- (2.50,2.48);

\draw[sliceline]
    (4.60,4.02) -- (4.60,2.48);

\draw[sliceline]
    (6.75,4.02) -- (6.75,2.48);

\draw[sliceline]
    (8.80,4.02) -- (8.80,2.48);

% =================================================
% BURST BOXES
% =================================================

\node[cburst] (b1) at (1.45,1.55) {
    \textbf{\textcolor{clientblue}{Burst 1}}\\[3pt]
    \textcolor{clientblue}
    {$\mathrm{C}\!\rightarrow\!\mathrm{S}$}\\[5pt]
    4 packets\\[2pt]
    Duration $T_1$
};

\node[sburst] (b2) at (3.55,1.55) {
    \textbf{\textcolor{serverred}{Burst 2}}\\[3pt]
    \textcolor{serverred}
    {$\mathrm{S}\!\rightarrow\!\mathrm{C}$}\\[5pt]
    3 packets\\[2pt]
    Duration $T_2$
};

\node[cburst] (b3) at (5.65,1.55) {
    \textbf{\textcolor{clientblue}{Burst 3}}\\[3pt]
    \textcolor{clientblue}
    {$\mathrm{C}\!\rightarrow\!\mathrm{S}$}\\[5pt]
    4 packets\\[2pt]
    Duration $T_3$
};

\node[sburst] (b4) at (7.75,1.55) {
    \textbf{\textcolor{serverred}{Burst 4}}\\[3pt]
    \textcolor{serverred}
    {$\mathrm{S}\!\rightarrow\!\mathrm{C}$}\\[5pt]
    6 packets\\[2pt]
    Duration $T_4$
};

\node[cburst] (b5) at (9.85,1.55) {
    \textbf{\textcolor{clientblue}{Burst 5}}\\[3pt]
    \textcolor{clientblue}
    {$\mathrm{C}\!\rightarrow\!\mathrm{S}$}\\[5pt]
    3 packets\\[2pt]
    Duration $T_5$
};

\end{tikzpicture}
    }
    % \caption{\textbf{Burst-level traffic representation.} Consecutive payload-bearing packets transmitted in the same direction form a burst, characterized by its direction, packet count, and duration.}
    \caption{\textbf{Burst-level traffic representation.} Consecutive packets transmitted in the same direction form a burst.}
    \label{fig:burst-representation}
    \vspace{-0.3cm}
\end{figure}
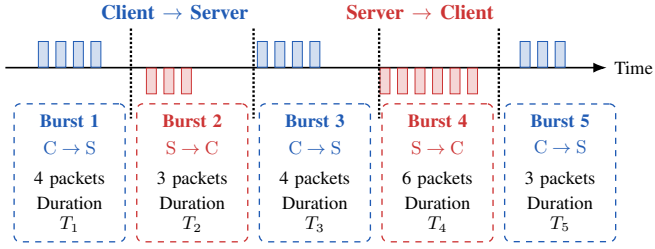

\subsubsection{LLMs Can Be Reliably Fingerprinted}
Since different models exhibit distinguishable traffic characteristics, we next examine which features contribute most to model fingerprinting. We group the 168 extracted features into six families by representation granularity: flow-level \emph{volume/direction} and \emph{rate}, packet-level \emph{packet size} and \emph{timing}, segment-level \emph{burst}, and sequence-level \emph{ordering}. Flow-level features summarize aggregate volume, directionality, and transmission rates, while packet-level features capture individual packet sizes and temporal relationships. Burst features characterize same-direction transmission segments, where a burst is a maximal sequence of consecutive payload-bearing packets transmitted in one direction, as illustrated in Fig.~\ref{fig:burst-representation}. These features summarize burst count, size, duration, and inter-burst spacing. Ordering features instead capture the global packet arrangement, including packet positions, early- and late-stage traffic concentration, response concentration, and sequence-order statistics.

\begin{tcolorbox}[
    colback=blue!5,
    colframe=blue!40!black,
    colbacktitle=blue!15!white,
    coltitle=black,
    fonttitle=\bfseries,
    title filled,
    boxrule=0.5pt,
    arc=2mm,
    left=0.0mm, right=0.1mm, top=0.1mm, bottom=0.01mm,
    title={\textit{\small Key Insight: Burst Features Dominate}}
]
\small Burst features alone identify the model nearly as well as all $168$ features ($94.2\%$ vs. $95.7\%$), showing that response packetization and streaming (not traffic volume) carry the signal. This lowers the cost of the attack, since coarse burst statistics are sufficient. It also implies that defenses hiding only a single feature such as packet size or timing leave the dominant side-channel open.
%Burst features alone capture nearly all information needed for model fingerprinting, showing that response packetization and streaming dominate the observable side channel.
\end{tcolorbox}

For each family, we perform two complementary experiments. In the
\emph{family-only} setting, the classifier is trained using only that family to
measure its predictive sufficiency. In the \emph{without-family} setting, the
family is removed and the classifier is retrained using all remaining features
to measure its unique contribution. We train a new RF for every
feature configuration and prompt-grouped fold using the same protocol as the
main evaluation, with preprocessing fitted only on the training partition.
Table~\ref{tab:model-feature-families} reports the mean and standard deviation
across the five test folds.

The feature families contain different numbers of features. Consequently,
family-only accuracy reflects the predictive sufficiency of the complete
family rather than a cardinality-controlled comparison. We therefore use the
\emph{without-family} results to assess each family's unique contribution.

\begin{table}[t]
    \centering
    \caption{\small Feature-family analysis for LLM-model fingerprinting using one repetition per model--prompt pair. Values are percentages (mean $\pm$ std dev across five prompt-grouped folds).}
    \label{tab:model-feature-families}
    \scriptsize
    \begin{tabular}{lccc}
        \toprule
        \textbf{Feature family} &
        \textbf{No. features} &
        \textbf{Family only (\%)} &
        \textbf{Without family (\%)} \\
        \midrule
        All features       & 168 & $95.67 \pm 0.70$ & --                  \\
        Burst              & 75  & $94.17 \pm 1.56$ & $86.33 \pm 2.67$   \\
        Timing             & 28  & $66.50 \pm 1.09$ & $95.67 \pm 1.49$    \\
        Packet size        & 21  & $87.83 \pm 1.62$ & $95.83 \pm 1.32$    \\
        Ordering           & 22  & $48.00 \pm 5.67$ & $95.83 \pm 0.83$   \\
        Volume/direction   & 10  & $30.17 \pm 4.87$ & $96.33 \pm 1.12$    \\
        Rate               & 12  & $47.83 \pm 4.51$ & $97.33 \pm 1.09$    \\
        \bottomrule
    \end{tabular}
    %\vspace{-0.25in}
\end{table}

\noindent\textbf{Burst alone could be enough.}  Using all $168$ features, the classifier identifies the deployed model with \(95.67\%\) accuracy, well above the \(10\%\) random-guessing baseline. Burst features achieve \(94.17\%\) in isolation, and removing them reduces accuracy by \(9.33\%\). A directional burst combines the volume and duration of consecutive packets traveling in one direction with their separation from adjacent transmissions. It therefore captures both the packetization and temporal structure of a streamed response. Packet-size and timing features are also informative in isolation, achieving
\(87.83\%\) and \(66.50\%\), respectively, but contribute little beyond burst
features because they capture overlapping information. In contrast, aggregate
volume and direction are much less effective because they discard the
packetization and temporal structure of the response.

\noindent\textbf{Why do these features work?} Two mechanisms explain why these features differ across models. First, autoregressive decoding time depends on the deployed model's architecture, parameter count, attention configuration, and execution characteristics. Prior work shows that these differences produce model-dependent inter-token timing patterns observable after responses are transmitted over a network~\cite{alhazbi2025rhythm}. Our evaluated models span approximately $3$B--$15$B parameters and several Llama, Phi, and Qwen architectures, providing substantial variation in the computation performed at each decoding step.

\begin{tcolorbox}[
    colback=blue!5,
    colframe=blue!40!black,
    colbacktitle=blue!15!white,
    coltitle=black,
    fonttitle=\bfseries,
    title filled,
    boxrule=0.5pt,
    arc=2mm,
    left=0.0mm, right=0.1mm, top=0.1mm, bottom=0.01mm,
    title={\textit{\small Key Insight: Fingerprints Reflect the Deployed LLM System}}
]
\small 
The fingerprint identifies the deployed serving system rather than architecture alone. Two post-trained models sharing the Qwen3-14B backbone remain separable, with $100\%$ and $96.7\%$ recall, while the most common confusions occur across different architectures. Thus, withholding model details does not prevent passive identification of serving variants.

%The fingerprint characterizes the deployed LLM serving system rather than the model architecture alone: two models sharing the same Qwen3-14B backbone remain separable after post-training ($100\%$ and $96.7\%$ recall), while the most frequent confusions occur between models with different architectures. The exposure therefore extends to post-trained variants, since withholding model details does not prevent a passive observer from identifying which serving system produced a response.
%The fingerprint characterizes the deployed LLM serving system rather than the model architecture alone. Models sharing the same pretrained model remain distinguishable after post-training, while different architectures can produce similar traffic, indicating that generation and serving behavior jointly determine the encrypted fingerprint.
\end{tcolorbox}

Second, training, post-training, alignment, and tokenization affect what each model generates for the same prompt. LLMmap demonstrates that identical queries elicit model-dependent response lengths, structures, wording, and refusal behavior~\cite{pasquini2025llmmap}. These textual differences become network-visible in our collection pipeline. The server streams each decoded text chunk to the client as a separate
Server-Sent Events (SSE) message over a persistent HTTP connection.
Consequently, differences in tokenization, word length, formatting, and
response structure alter the sizes and sequence of transmitted chunks, while differences in decoding speed alter their emission intervals.

Architecture alone, however, does not explain the result. Qwen3-14B-Base and
MegaScience-14B share the same Qwen3-14B backbone, but the latter is further
trained for scientific reasoning; the classifier obtains \(100\%\) and
\(96.7\%\) recall for these models, respectively. This indicates that
post-training-induced response behavior can change the traffic fingerprint
without changing the base architecture. Conversely, most classification errors
occur between Phi-4-Mini and LlamaTwin-8B, showing that models with different
architectures can still produce overlapping traffic behavior. We therefore interpret the measured signature as a fingerprint of the deployed
LLM serving system rather than the model architecture alone. The fingerprint
reflects the combined effects of the model architecture, post-training,
tokenization, and generation behavior.

% Packet-size features are highly sufficient, achieving \(87.83\%\) alone, but
% their removal causes no measurable degradation because burst-byte features
% retain overlapping information from the same packet sequence. Timing features
% similarly achieve \(66.50\%\) alone, while burst duration, burst separation,
% and rate features preserve related temporal information. Aggregate
% volume/direction is considerably weaker because it collapses the response into
% a few totals and discards the packetization and temporal structure that
% distinguishes the model-serving pipelines. 

\subsubsection{Prompt-category fingerprinting}
We next investigate whether encrypted traffic reveals the semantic category of
a prompt when the deployed LLM is fixed. We train a separate six-class Random
Forest classifier for each of the ten LLMs, preventing model-specific traffic
differences from directly determining the predicted category. We use all available traces and five common prompt-grouped folds. In
each fold, the classifier for a given model is trained on all repetitions of
48 prompts and tested on all repetitions of 12 completely unseen prompts.
All repetitions of a prompt remain in the same fold. We first average accuracy
across the five folds for each model and then report the mean and standard
deviation across the ten models. Random guessing obtains \(16.67\%\).

\begin{table}[t]
\centering
\caption{\small Prompt-category fingerprinting using different traffic views and response-only feature families. ``Only'' uses only the indicated response-feature family, whereas ``Without'' removes it from the complete 22-feature response representation. Values are mean accuracy across the ten LLM-specific classifiers.}
\label{tab:category-response-families}
\scriptsize
\begin{tabular}{lcccc}
\toprule
\textbf{Features} & \textbf{No.} & \textbf{Only (\%)} & \textbf{Without (\%)} & \textbf{Drop (pp)} \\
\midrule

\multicolumn{5}{l}{\textit{Traffic-view sanity check}}\\
All traffic        & 168 & $65.10 \pm 7.93$ & -- & -- \\
Request only       & 16  & $51.68 \pm 0.81$ & -- & -- \\
Response only      & 22  & $67.37 \pm 11.24$ & -- & -- \\

\midrule
\multicolumn{5}{l}{\textit{Response-only feature families}}\\

Volume             & 2 & $37.12 \pm 7.08$ &
$67.32 \pm 11.37$ & $0.05$ \\
Packet size        & 7 & $49.11 \pm 9.16$ &
$59.03 \pm 8.82$ & $8.34$ \\
Timing             & 8 & $55.40 \pm 9.44$ &
$60.11 \pm 10.19$ & $7.26$ \\
Rate               & 2 & $42.23 \pm 7.99$ &
$67.37 \pm 11.29$ & $0.01$ \\
Temporal ordering  & 3 & $36.63 \pm 5.78$ &
$65.63 \pm 12.02$ & $1.74$ \\

\midrule
Random guessing    & -- & $16.67$ & -- & -- \\
\bottomrule
\end{tabular}
%\vspace{-0.2in}
\end{table}

\noindent\textbf{Response-side prompt-category fingerprinting.}
Prompt-category classification can be biased by encrypted request traffic because categories differ in prompt length. Although each prompt fits within one request packet, median encrypted sizes vary widely across C1--C6: 1,388, 131, 77.5, 146.5, 283, and 95.5 bytes. Accordingly, Table~\ref{tab:category-response-families} shows that request traffic alone achieves 51.68\% accuracy, revealing prompt length as a strong shortcut. Because prompt length can be altered without changing task semantics, we focus on the more challenging response-only setting using 22 LLM-to-user features.

% Prompt-category classification can be strongly influenced by encrypted request
% traffic because prompts from different categories have different lengths. In
% our dataset, each prompt fits within a single request packet, but the encrypted
% request size still varies substantially across categories; the median sizes for
% C1--C6 are 1,388, 131, 77.5, 146.5, 283, and 95.5 bytes, respectively.
% Accordingly, Table~\ref{tab:category-response-families} shows that request
% traffic alone achieves \(51.68\%\) accuracy, indicating that prompt length
% provides a strong shortcut for category fingerprinting. Since an adversary can
% easily modify prompt length without changing the task semantics, we instead
% consider the more challenging response-only setting using 22 features computed
% from LLM-to-user traffic.

\begin{tcolorbox}[
    colback=blue!5,
    colframe=blue!40!black,
    colbacktitle=blue!15!white,
    coltitle=black,
    fonttitle=\bfseries,
    title filled,
    boxrule=0.5pt,
    arc=2mm,
    left=0.0mm, right=0.1mm, top=0.1mm, bottom=0.01mm,
    title={\textit{\small Key Insight: Categories Leak Through Generation Behavior}}
]
\small With request packets excluded and all test prompts held out, response packetization and timing alone identify the prompt category with 67.4\% accuracy, compared with a 16.7\% baseline. This reveals the type of task a user is performing even when prompt length changes, because the leakage originates from the generated response rather than the request.

%With request packets excluded and test prompts fully held out, response packetization and timing alone recover the prompt category at $67.4\%$ against a $16.7\%$ baseline. The privacy risk is direct, because an adversary learns what kind of task a user is performing even if prompt length is altered, as the leakage arises from the generated response itself.
%Prompt semantics survive encryption because different categories induce different response structures and generation behaviors. These differences manifest as distinct encrypted packet sizes and transmission timing.
\end{tcolorbox}

Using only response traffic, Table~\ref{tab:category-response-families}
achieves \(67.37\%\) accuracy, exceeding both the request-only
(\(51.68\%\)) and all-traffic (\(65.10\%\)) settings. Timing is the strongest
feature family in isolation (\(55.40\%\)), while packet size contributes the
most unique information, reducing accuracy by \(8.34\) percentage points when
removed, followed by timing (\(7.26\) points). Removing volume or rate causes
almost no degradation, indicating that they are largely redundant with the
more detailed packet-size and timing statistics. Because request-side features
are excluded and complete prompts are held out, the observed \(67.37\%\)
accuracy cannot be explained by prompt length. Instead, different prompt
categories induce different response structures and decoding behaviors that, in
our SSE-based serving pipeline, translate into distinct encrypted chunk sizes
and transmission intervals, consistent with prior
work~\cite{pasquini2025llmmap,weiss2024your}. Overall, prompt-category leakage
is driven primarily by response packetization and generation timing rather than
aggregate response volume.

\begin{table}[t]
\centering
\caption{\small Per-category performance using response-only traffic and all
repetitions. Metrics are computed from pooled out-of-fold predictions. The
final column reports the most frequent wrong prediction as a percentage of the
true category's traces.}
\label{tab:category-class-performance}
\scriptsize
\setlength{\tabcolsep}{10.5pt}
\begin{tabular}{@{}lcccl@{}}
\toprule
\textbf{Category} & \textbf{Precision} & \textbf{Recall} &
\textbf{F1} & \textbf{Main confusion} \\
\midrule
C1 & 75.65\% & 79.91\% & 77.72\% & C4 (7.92\%) \\
C2 & 58.31\% & 52.82\% & 55.43\% & C3 (16.21\%) \\
C3 & 66.74\% & 67.10\% & 66.92\% & C2 (15.07\%) \\
C4 & 69.07\% & 73.15\% & 71.05\% & C6 (9.14\%) \\
C5 & 66.08\% & 59.57\% & 62.66\% & C6 (12.11\%) \\
C6 & 66.37\% & 71.15\% & 68.68\% & C4 (8.72\%) \\
\bottomrule
\end{tabular}\vspace{-0.2in}
\end{table}

Table~\ref{tab:category-class-performance} reports the per-category precision,
recall, F1 score, and most frequent misclassification using response-only
traffic. C2 and C3 are most frequently confused, because code
generation and mathematical reasoning often produce similarly structured
multi-step outputs. To investigate this, we compare category distributions
using the 22 response-only features. Each model--prompt pair is represented by the median over its repetitions, robustly normalized within each model, and compared using the Wasserstein distance. Among all 15 category pairs, C2 and C3 have the smallest average distance (\(0.786\)), confirming that their responses are the most similar. This similarity is driven by timing: they have the closest mean response inter-arrival time and the second-closest median inter-arrival time, packet rate, and byte rate, whereas response length, byte volume, and duration differs substantially.

% \noindent\textbf{Insight (maybe for all the section)}
% Taken together, the experiments show that timing carries predictive information
% for both attacks, but its contribution differs between them. For model
% fingerprinting, timing features are informative in isolation but largely
% redundant with burst-duration, burst-separation, and rate measurements. For
% prompt-category fingerprinting with the model fixed, removing response timing
% reduces accuracy by \(7.26\%\), showing that it contributes
% information not fully captured by packet size. These measurements establish
% statistical associations rather than a specific causal mechanism: the observed
% packet intervals jointly reflect model generation, serving and buffering
% behavior, packetization, and the network path, which our traces do not separate.

\begin{figure}[t]
    \centering
    \includegraphics[width=\columnwidth]{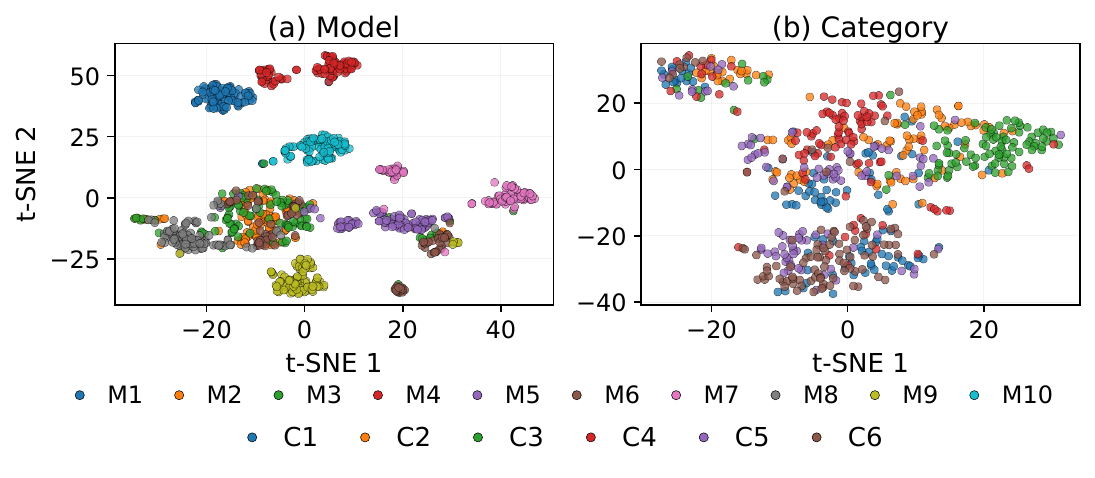}\vspace{-0.32cm}
    \caption{\small t-SNE visualization of encrypted user--LLM traffic. Panel (a) compares ten LLMs
    executing the same prompt using all 168 features. Panel (b)
    compares six prompt categories for a fixed LLM using all 22
    response-side features. Each point represents one traffic trace; colors denote
    models or prompt categories.}
    \label{fig:user-llm-tsne-comparison}\vspace{-0.2in}
\end{figure}

\subsubsection{Feature-space Visualization}
Figure~\ref{fig:user-llm-tsne-comparison} provides a qualitative visualization of the feature spaces for model and prompt-category fingerprinting. The model embedding exhibits clearer local separation than the prompt-category embedding, consistent with the corresponding classification accuracies ($95.67\%$ and $67.37\%$). Because t-SNE preserves local neighborhoods but distorts global distances, it is intended only as a qualitative illustration rather than an objective measure of performance.

\subsubsection{Information Leakage Is Concentrated in a Few Feature Families}
\label{sec:information-leakage}
The preceding experiments assess LLM fingerprinting using Random Forest classification. As accuracy depends on the chosen classifier and does not directly quantify the information revealed about LLM identity, we complement it with an information-theoretic analysis. Following~\cite{li2018measuring}, we measure leakage as the mutual information between the observed traffic features (X) and LLM identity (M).

% The preceding experiments evaluate LLM fingerprinting through the
% classification performance of a Random Forest classifier. Classification
% accuracy, however, is classifier dependent and does not directly quantify how
% much information the encrypted traffic reveals about the LLM identity. We
% therefore complement the classification results with an information-theoretic
% analysis. Specifically, we measure the information leakage, defined as the
% mutual information between the observed traffic features \(X\) and the LLM
% identity \(M\), following the methodology of~\cite{li2018measuring}.

We retain 59 response-side features and estimate their mutual information with model identity using WeFDE~\cite{li2018measuring}. Under a uniform prior over ten models, the maximum leakage is ($H(M)=\log_2 10=3.32$) bits. Within each of five prompt-grouped folds, we estimate per-feature leakage using adaptive kernel density estimation and 5,000 Monte Carlo samples, remove features with normalized mutual information above 0.9, and combine the remaining class-conditional marginal likelihoods under a conditional-independence approximation. All estimation, feature selection, and density fitting use only the training prompts.

% We retain $59$ response-side features and estimate their mutual information
% with model identity using the WeFDE methodology~\cite{li2018measuring}.
% Assuming a uniform prior over the ten models, the maximum possible leakage is
% $H(M)=\log_2 10 = 3.32$ bits. Within each of five prompt-grouped folds, we estimate individual-feature
% leakage using adaptive kernel density estimation and $5,000$ Monte Carlo samples,
% remove features whose normalized mutual information exceeds $0.9$, and combine
% the remaining class-conditional marginal likelihoods. This provides a scalable
% conditional-independence approximation of WeFDE. All estimation, feature
% selection, and density fitting use only the training prompts.

\begin{tcolorbox}[
    colback=blue!5,
    colframe=blue!40!black,
    colbacktitle=blue!15!white,
    coltitle=black,
    fonttitle=\bfseries,
    title filled,
    boxrule=0.5pt,
    arc=2mm,
    left=0.0mm, right=0.1mm, top=0.1mm, bottom=0.01mm,
    title={\textit{\small Key Insight: Information-Theoretic Analysis Confirms Burst Dominance}}
]
\small Classifier-free analysis confirms the ranking: burst-byte features leak 3.02 of the maximum 3.32 bits about model identity and alone support $86.9\%$ accuracy, while total response volume yields only $23.1\%$. Because the leakage is a property of the encrypted traffic rather than of any particular classifier, the risk persists and can only be reduced by reshaping the traffic profile itself.
%Information-theoretic analysis confirms the classifier-based result: burst-byte features leak \(3.02\) of the maximum \(3.32\) bits about model identity.
\end{tcolorbox}

We relate leakage to classification using Fano’s inequality~\cite{fano1961transmission},
\begin{equation}
H(M\mid X)\leq h_2(P_e)+P_e\log_2(9).
\end{equation}
Solving for ($P_e$) yields a lower bound on classification error and, equivalently, an upper bound ($1-P_e^{\mathrm{Fano}}$) on accuracy. Because this bound is generally loose and not Bayes-optimal, we compare it with a Random Forest trained on the same selected features and evaluated on held-out prompts.

%This bound is generally loose and does not estimate Bayes-optimal accuracy, we compare it with a Random Forest trained on the same selected features and evaluated on held-out prompts.

% We relate leakage to classification through Fano's
% inequality~\cite{fano1961transmission},
% \begin{equation}
%  H(M\mid X)\leq h_2(P_e)+P_e\log_2(9).
% \end{equation}
% Solving this inequality gives an error lower bound
% \(P_e^{\mathrm{Fano}}\), and therefore an accuracy upper bound
% \(1-P_e^{\mathrm{Fano}}\). This bound is generally loose and is not an estimate
% of Bayes-optimal accuracy. We compare it with a Random Forest trained on the
% same selected features and evaluated on held-out prompts.

\begin{figure}[t]
    \centering
    \includegraphics[width=\columnwidth]
{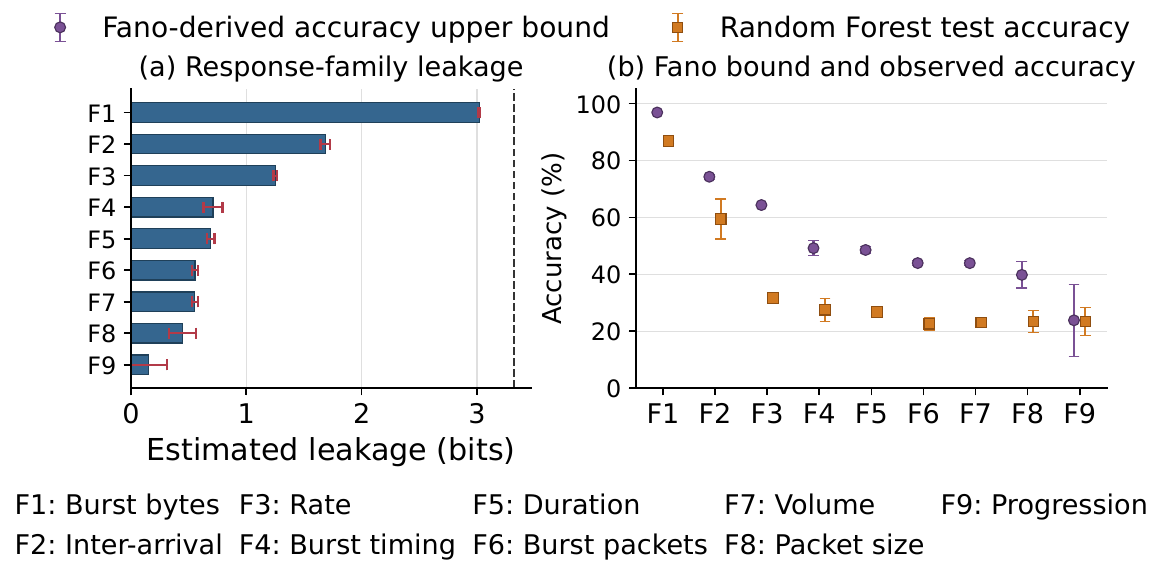}\vspace{-0.1cm}
    \caption{\small Response-side model leakage and fingerprinting performance.
    Families are ordered by estimated leakage, with F1--F9 defined in the legend. (a) Response-family leakage \(I(X;M)\); the dashed line marks the maximum
possible leakage \(H(M)=3.32\) bits. (b) Fano-derived accuracy upper bound and held-out
    Random Forest test accuracy. Error bars denote 95\% confidence intervals
    across five prompt-grouped folds.}
    \label{fig:response-leakage-fano}\vspace{-0.1in}
\end{figure}

Figure~\ref{fig:response-leakage-fano} shows that response burst-byte features leak (3.02) bits and achieve 86.85\% test accuracy, compared with 23.08\% using total response volume alone. While total volume varies with the prompt, burst statistics capture how encrypted responses are segmented into application exchanges. Notably, the minimum response-burst size has zero within-model interquartile range and assumes eight distinct values across the ten models, consistent with model-dependent streaming, termination, or metadata records in the serving stack. Because payloads are encrypted, we interpret this as a deployed-system fingerprint rather than evidence that model architecture alone determines packetization.

% Figure~\ref{fig:response-leakage-fano} shows that response burst-byte features
% leak \(3.02\) bits and achieve \(86.85\%\) test accuracy, whereas total response
% volume achieves only \(23.08\%\). Total volume varies substantially with the
% prompt, while burst statistics retain how the encrypted response is segmented
% into application exchanges. In particular, the minimum response-burst size has
% zero within-model interquartile range and takes eight distinct values across
% the ten models. This stable pattern is consistent with model-dependent
% streaming, termination, or metadata records produced by the serving stack.
% Because their contents are encrypted, we interpret this as a deployed-system
% fingerprint rather than evidence that model architecture alone determines
% packetization.

Response inter-arrival features rank second, leaking 1.68 bits and achieving 59.4\% accuracy. Their model-level medians range from approximately 19 to 52 ms, reflecting differences in generation and stream-delivery intervals, although overlapping distributions limit classification. Unlike burst-byte features, which exhibit near-zero within-model variance and behave almost categorically, inter-arrival timing varies continuously with decoding speed and network delivery. Rate features leak 1.25 bits but achieve only 31.65\% accuracy, while packet-size features reach 23.35\% because they discard burst structure and overlap under common transport segmentation. The gap between the Fano bound and observed accuracy reflects bound looseness and finite-sample leakage estimation, not distance from Bayes-optimal performance.

% Response inter-arrival features rank second, leaking \(1.68\) bits and
% achieving \(59.42\%\) accuracy. Their model-level medians range from
% approximately \(19\) to \(52\) ms, reflecting different generation and
% stream-delivery intervals, although overlap between models limits accuracy.
% Unlike burst-byte features, whose near-zero within-model variance makes them
% almost categorical, inter-arrival timing varies continuously with decoding
% speed and network delivery, leading to overlapping model distributions. Rate
% features leak \(1.25\) bits but achieve only \(31.65\%\) accuracy, while
% packet-size features achieve \(23.35\%\) because they discard burst structure
% and overlap under common transport segmentation. The gap between the Fano
% bound and observed accuracy reflects the looseness of the bound and finite-
% sample leakage estimation rather than distance from Bayes-optimal
% performance.

\subsection{Encrypted Traffic Reveals Multi-Agent Tasks}
\label{sec:multi-agent-results}

\begin{tcolorbox}[
    colback=blue!5,
    colframe=blue!40!black,
    colbacktitle=blue!15!white,
    coltitle=black,
    fonttitle=\bfseries,
    title filled,
    boxrule=0.5pt,
    arc=2mm,
    left=0.0mm, right=0.1mm, top=0.1mm, bottom=0.01mm,
    title={\textit{\small Key Insight: Workflow Alone Leaks Task Identity}}
]
\small Even after removing all packet-size, byte-volume, and burst-byte features, workflow features alone identify the collaborative task at 54.0\% (graph) and 58.7\% (star) balanced accuracy against the 10\% chance level, rising to 85.3\% and 86.0\% with all 247 features. For deployments of collaborative multi-agent systems, coordination structure is itself a sensitive attribute, so packet-size padding alone would not conceal what agents are doing.
%Even after removing all packet-size and byte-volume information, workflow features alone identify multi-agent tasks, showing that coordination structure itself leaks task identity.
\end{tcolorbox}

We next evaluate whether encrypted agent--LLM traffic is sufficient to identify
the task performed by a collaborative multi-agent system. We first
characterize the traffic generated by different tasks and then measure
fingerprinting performance under different communication topologies. Finally,
we evaluate the attack when the observer captures encrypted traffic from only a
subset of the agent--LLM communication channels, rather than all agents
participating in the execution.

\paragraphB{Tasks Produce Distinct Workflow Signatures}
We next rank features using Random Forest mean decrease in impurity, computed
only on the training partitions of five task-grouped folds. Table~\ref{tab:marble-feature-ranks} and Fig.~\ref{fig:marble-feature-importance}(a) show that $8$ of the $10$ highest-ranked features are shared by both graph and star topologies. Burst volume, packet size, and early outgoing traffic dominate both,
indicating that task identity is reflected consistently in response packetization and execution structure.

% We examine whether task categories produce distinguishable encrypted
% traffic under the graph topology. To avoid overweighting tasks with more
% successful repetitions, we compute each feature's median across repetitions
% and then across the $15$ tasks in each category. Fig.~\ref{fig:marble-traffic-characteristics}
% shows representative standardized features.

% \begin{figure}[t]
%     \centering
%     \includegraphics[width=0.35\textwidth]
%     {img/marble_10category_traffic_heatmap.pdf}\vspace{-0.1in}
%     \caption{\small Encrypted-traffic characteristics of $10$ MARBLE task categories under the graph topology. Each task contributes once after aggregation across repeated executions; cells report category medians.}
%     \label{fig:marble-traffic-characteristics}\vspace{-0.1in}
% \end{figure}
% The categories differ in combinations of volume, duration, rate, bursts, and
% idle behavior rather than in traffic volume alone. For example, bargaining
% generates long, high-volume executions with relatively low packet rates,
% whereas bug-fixing and debate produce less traffic at higher rates. Coding
% exhibits longer idle periods, consistent with fewer but longer generation
% steps. These patterns reflect differences in coordination rounds, call
% duration, and response structure, producing task-dependent signatures even
% when payload is encrypted.

\begin{table}[t]
\centering
\caption{\small Top-ranked task-fingerprinting features. The 75th percentile is
denoted by $Q_{75}$; outgoing and incoming indicate agent-to-LLM and
LLM-to-agent traffic.}
\label{tab:marble-feature-ranks}
\scriptsize\vspace{-0.05in}
\begin{tabular}{@{}c |p{0.39\columnwidth} |p{0.39\columnwidth}@{}}
\toprule
\textbf{Rank} & \textbf{Graph} & \textbf{Star} \\
\midrule
1  & Outgoing burst bytes, $Q_{75}$ & Outgoing burst bytes, $Q_{75}$ \\
2  & Outgoing packet size, std. & Burst bytes, median \\
3  & First-30 outgoing bytes & Outgoing packet size, mean \\
4  & Incoming burst bytes, median & Incoming burst bytes, median \\
5  & Burst bytes, $Q_{75}$ & Outgoing packet size, $Q_{75}$ \\
6  & Outgoing burst bytes, std. & Outgoing packet size, std. \\
7  & Outgoing packet size, mean & Packet size, $Q_{75}$ \\
8  & Burst bytes, median & First-30 outgoing bytes \\
9  & Outgoing packet size, $Q_{75}$ & Burst bytes, $Q_{75}$ \\
10 & Outgoing burst bytes, maximum & Activity-episode bytes, minimum \\
\bottomrule
\end{tabular}\vspace{-0.1in}
\end{table}

To isolate workflow structure from response size, we exclude packet-size, byte-volume, and burst-byte features, retaining 134 timing, burst, idle, directional, and progression features. The top 100 workflow features achieve 54.0\% balanced accuracy under graph topology and 58.7\% under star, well above the 10\% chance level. Figure~\ref{fig:marble-feature-importance}(b) shows partial category overlap, reflecting shared plan--execute--review workflows across tasks. Using all 247 features raises balanced accuracy to 85.3\% and 86.0\%, with macro-F1 scores of 84.6\% and 85.2\%, respectively. Thus, workflow structure alone reveals task identity, while response volume and packetization provide substantial complementary leakage.

% To separate workflow structure from response size, we exclude packet-size,
% byte-volume, and burst-byte features and retain $134$ timing, burst, idle,
% directional, and progression features. The top $100$ workflow features achieve
% $54.0\%$ balanced accuracy for graph and $58.7\%$ for star, well above the 10\%
% chance level. Figure~\ref{fig:marble-feature-importance}(b) provides a
% qualitative view; although several categories overlap
% because different tasks can share similar plan--execute--review workflows. Using all $247$ features increases balanced accuracy to $85.3\%$ for graph and
% $86.0\%$ for star, with macro-F1 scores of $84.6\%$ and $85.2\%$. Thus, workflow
% organization alone reveals task identity, while response volume and
% packetization provide substantial complementary information.

% \paragraph{Fingerprinting accuracy.}
% Figure~\ref{fig:marble-fingerprinting}(a) shows that aggregate volume already
% reveals substantial task information, obtaining 71.3\% balanced accuracy for
% graph and 66.7\% for star. All engineered features increase accuracy to 85.3\%
% and 86.0\%, with corresponding macro-F1 scores of 84.6\% and 85.2\%. Selecting
% the 20 highest-ranked features within each training fold obtains 84.0\% for
% graph and 90.7\% for star. The improvement over volume-only classification
% shows that the attack also exploits timing, packet-size, burst, and execution-
% progress structure.

\paragraphB{Task fingerprints persist under topology changes and partial visibility}
We evaluate whether task fingerprints transfer across agent-coordination
topologies and remain effective when only a subset of agent traffic is
observable. 

\begin{tcolorbox}[
    colback=blue!5,
    colframe=blue!40!black,
    colbacktitle=blue!15!white,
    coltitle=black,
    fonttitle=\bfseries,
    title filled,
    boxrule=0.5pt,
    arc=2mm,
    left=0.0mm, right=0.1mm, top=0.1mm, bottom=0.01mm,
    title={\textit{\small Key Insight: Partial Visibility Still Leaks Tasks}}
]
\small 
From a single agent’s traffic, an adversary infers the collaborative task with 66.5\% balanced accuracy under graph topology and 78\% under star topology, increasing to 84\% and 86\% with visibility into up to four agents. Cross-topology accuracy drops from 83\% to 52\% but remains well above random guessing. Thus, partial agent--LLM traffic visibility poses substantial privacy risk, while topology changes reduce but do not eliminate the leakage.

%An adversary observing traffic from a single agent infers the collaborative task at 66.5\% (graph) and 78.0\% (star) balanced accuracy, rising to 84.3\% and 85.9\% when traffic from up to four agents is visible, while cross-topology accuracy drops to 52.0\% vs. 83.0\% within topology yet stays far above the random-guessing baseline. Even a weak adversary observing only a subset of agent–LLM communications therefore poses a substantial risk, and changing the coordination topology reduces but does not remove the exposure.
% Multi-agent task fingerprints remain effective even with incomplete observations. A single agent already reveals substantial task information, while additional agents provide complementary workflow signals.
\end{tcolorbox}

For topology transfer, we use feature sets of $3$, $6$, $10$, $50$, $100$,
and all $247$ features. For each size below $247$, we sample $20$ label-independent feature subsets and train a new Random Forest for every subset and task-grouped fold. Within-topology evaluation trains and tests on the same topology, while cross-topology evaluation trains on graph and tests on star, or vice versa. For partial visibility, we construct nested traces containing traffic from up
to \(k\) agents. We sample ten agent orderings per execution and retrain the
classifier for every topology, visibility level, ordering, and fold. Agent
identities are used only to construct the restricted traces and are never
provided to the classifier.

\begin{figure}[t]
    \centering
    \includegraphics[width=\columnwidth]
    {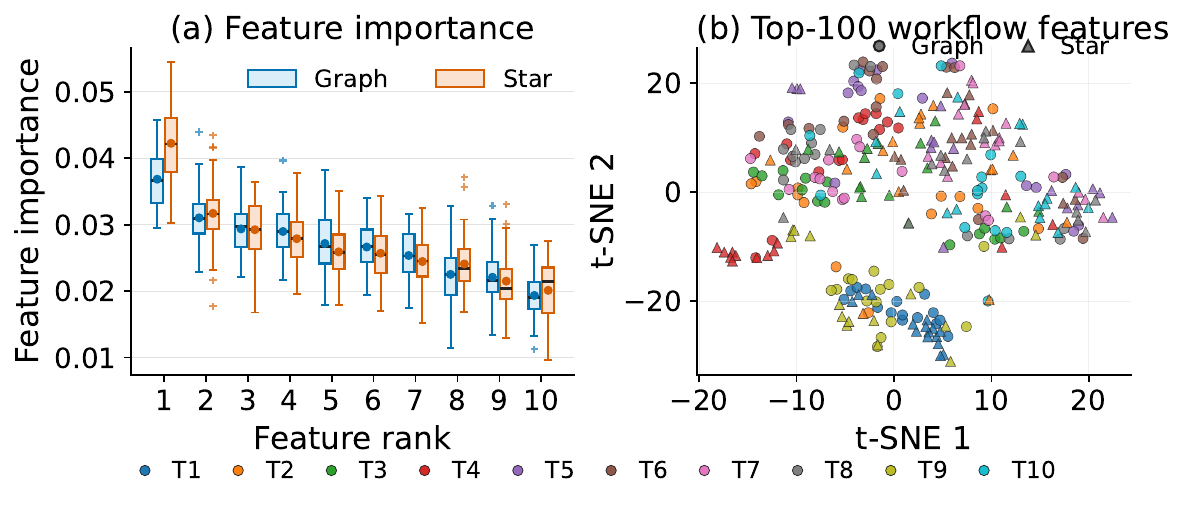}\vspace{-0.05in}
    \caption{\small Multi-agent task-fingerprinting features.
    (a) Importance of the $10$ highest-ranked features under graph and star
    topologies. (b) Joint t-SNE using the $100$ highest-ranked
    workflow-only features. Colors denote task categories; marker shapes
    denote topology.}
    \label{fig:marble-feature-importance}\vspace{-0.1in}
\end{figure}

Figure~\ref{fig:topology-transfer-visibility}(a) shows that additional features
improve task fingerprinting in both settings. With all features,
within-topology accuracy reaches $83.0\%$, whereas cross-topology accuracy
reaches only $52.0\%$. This transfer gap indicates that the fingerprint depends
partly on the agents' coordination structure. A classifier trained on one
topology therefore does not observe exactly the same traffic signature under
another. The gap likely arises because topology shapes the observable traffic:
the star topology routes coordination through a central hub, whereas the graph
topology distributes it across peer-to-peer links, causing the classifier to
learn topology-specific as well as task-dependent patterns. Figure~\ref{fig:topology-transfer-visibility}(b) shows that restricted traffic
visibility does not eliminate task leakage. Observing one agent yields $66.5\%$
accuracy under graph and $78.0\%$ under star. Accuracy increases to $84.3\%$ and
$85.9\%$, respectively, when traffic from up to four agents is visible. Thus, an
individual agent's interactions already reveal a substantial portion of the
task signature, while additional agents expose complementary workflow
information and strengthen the attack.

\subsection{Fingerprinting Persists Under Realistic Conditions}

\begin{tcolorbox}[
    colback=blue!5,
    colframe=blue!40!black,
    colbacktitle=blue!15!white,
    coltitle=black,
    fonttitle=\bfseries,
    title filled,
    boxrule=0.5pt,
    arc=2mm,
    left=0.0mm, right=0.1mm, top=0.1mm, bottom=0.01mm,
    title={\textit{\small Key Insight: Lightweight Countermeasures Reduce but Do Not Eliminate Fingerprinting}}
]
\small 
Prompt reformulation reduces model-fingerprinting accuracy by 24\% and temperature randomization by up to 18\%, but performance remains well above the 10\% random baseline. Expanding the candidate set from 5 to 20 LLMs lowers accuracy only from 98.5\% to 92\%. Thus, from a risk perspective, lightweight defenses reduce but do not eliminate leakage risk, and model fingerprints remain highly discriminative even at larger scale.

%Prompt reformulation reduces model-fingerprinting accuracy by 23.7\% and decoding-temperature randomization by up to 18.2\%, yet accuracy remains well above the 10\% random baseline; meanwhile, growing the candidate set from 5 to 20 LLMs lowers accuracy only from 98.5\% to 92.2\%. In risk terms, lightweight application- and serving-level defenses reduce but do not neutralize the exposure, and the fingerprints remain discriminative as the candidate set grows.
% Prompt reformulation and decoding variation reduce fingerprinting accuracy, but substantial leakage remains. These lightweight defenses are promising risk-reduction mechanisms that require further investigation.
\end{tcolorbox}

\subsubsection{Lightweight Countermeasures}

We evaluate two lightweight countermeasures that require no network-protocol changes. Prompt reformulation rewrites or wraps the user’s request while preserving its semantics; we use the LLMmap strategy, which inserts execution-trigger instructions before and after the original prompt. Decoding variation instead changes the temperature across requests rather than using a fixed setting. As shown in Figure~\ref{fig:fingerprinting-realistic}(a), prompt reformulation reduces model-fingerprinting accuracy from 93.2\% to 69.5\%, while temperatures of 0.3, 0.5, and 1.5 yield 75.0\%, 88.8\%, and 80.0\%, respectively. Both countermeasures weaken fingerprints learned under a fixed configuration, but accuracy remains well above the 10\% random baseline.

% We evaluate two lightweight countermeasures that require no changes to the
% network protocol. The first is prompt reformulation, where the application
% rewrites or wraps the user's request while preserving its semantics. We adopt
% the LLMmap prompt-crafting strategy, which inserts execution-trigger
% instructions before and after the original prompt. The second is decoding
% variation, where the service changes the decoding temperature across requests
% instead of using a fixed configuration. Figure~\ref{fig:fingerprinting-realistic}(a) shows that prompt reformulation
% reduces model-fingerprinting accuracy from $93.2\%$ to $69.5\%$. Varying the
% decoding temperature also weakens fingerprinting, yielding accuracies of
% $75.0\%$, $88.8\%$, and $80.0\%$ at temperatures 0.3, 0.5, and 1.5, respectively. Both approaches reduce the effectiveness of fingerprints learned under a fixed configuration, yet accuracy remains well above the 10\% random baseline.

% These results suggest that prompt reformulation and decoding variation are useful risk-reduction mechanisms rather than complete defenses.

\subsubsection{Scalability}

We next evaluate model fingerprinting as the candidate set grows from $5$ to $20$
LLMs. Figure~\ref{fig:fingerprinting-realistic}(b) compares our 168-feature
representation with a 35-feature baseline using a matched experimental
protocol. The proposed representation maintains high accuracy as the number of
candidate models increases, decreasing from $98.5\%$ with five models to $92.2\%$
with twenty models. In contrast, the 35-feature baseline drops from $76.0\%$ to
$48.8\%$, widening the performance gap as more models are added. This indicates
that modeling the structural characteristics of encrypted traffic, rather than
only aggregate timing and packet-size statistics, produces fingerprints that
remain discriminative as the candidate set grows.

\begin{figure}[t]
    \centering
    \includegraphics[width=0.49\textwidth]
    {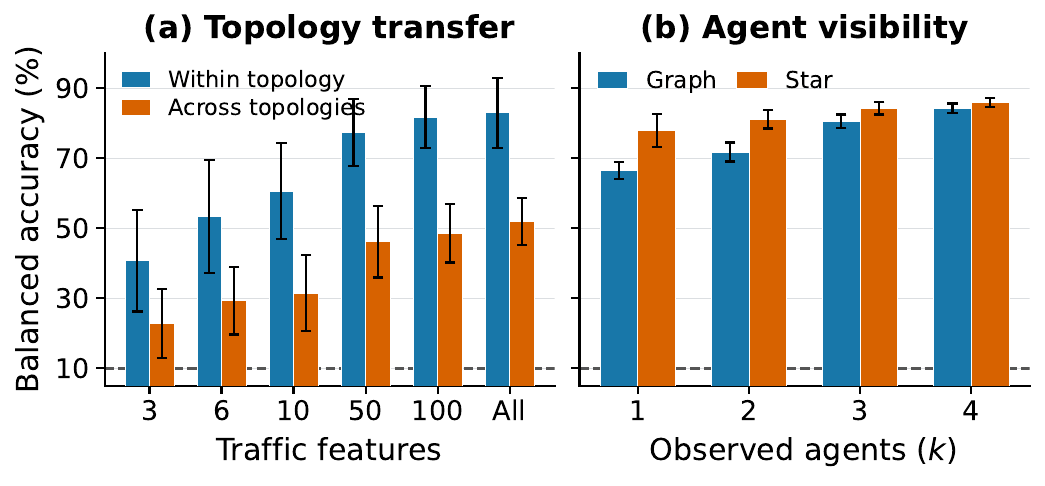}\vspace{-0.1cm}
    \caption{\small Multi-agent task fingerprinting under topology changes and partial
    traffic visibility. (a) Within-topology and cross-topology accuracy as the
    number of traffic features increases. (b) Accuracy when the attacker
    observes traffic from up to \(k\) agents. Error bars summarize variation
    across feature subsets and agent selections, respectively. The dashed line
    denotes the 10\% random-guessing baseline.}
    \label{fig:topology-transfer-visibility}\vspace{-0.1in}
\end{figure}

%\vspace{-0.1in}

% \noindent\textbf{Insight.}
% Fingerprinting remains practical under more realistic deployment conditions.
% Prompt reformulation and decoding variation reduce the effectiveness of
% fingerprints learned under a fixed configuration but do not eliminate them,
% while the proposed 168-feature representation maintains high accuracy even as
% the candidate set grows to 20 LLMs. Together, these results indicate that the
% observed fingerprints capture stable structural properties of encrypted
% user--LLM interactions rather than artifacts of a particular prompt or
% experimental setting.

\section{Discussion}
\label{sec:discuss}

\subsection{Potential Countermeasures}
\label{sec:countermeasure}

Countermeasures against encrypted LLM-traffic fingerprinting can be deployed at
different stages of the serving pipeline. Some are specific to LLM
applications and serving systems, whereas others can be adapted from the
traffic-analysis and website-fingerprinting literature.

\noindent\textbf{Application- and model-serving defenses.}
Applications can reformulate prompts through rewriting or paraphrasing while preserving the intended task. Using LLMmap’s prompt-crafting strategy reduces model-fingerprinting accuracy from 93.2\% to 69.5\%, showing that semantically equivalent prompts can substantially alter encrypted traffic. Providers can likewise vary decoding parameters such as temperature, top-p, top-k, or repetition penalties. Temperature variation also reduces accuracy, though non-monotonically, indicating that its effectiveness depends on the prompt and generated response. Overall, application- and model-level adaptations reduce but do not eliminate traffic fingerprintability.

% Applications can reformulate user prompts while preserving their intended task,
% for example through prompt rewriting or paraphrasing. Our evaluation shows that
% the prompt-crafting strategy of LLMmap reduces model-fingerprinting accuracy
% from 93.2\% to 69.5\%, demonstrating that semantically equivalent prompt
% reformulations substantially alter encrypted traffic. Similarly, providers can
% vary decoding parameters such as temperature, top-$p$, top-$k$, or repetition
% penalties. Our experiments show that changing the decoding temperature also
% reduces fingerprinting accuracy, although the effect is not monotonic,
% indicating that its benefit depends on the prompt and generated output.
% Together, these results suggest that application- and model-level adaptations
% can reduce, but not eliminate, traffic fingerprintability.

\begin{figure}[!t]
    \centering
    \includegraphics[width=0.5\textwidth]{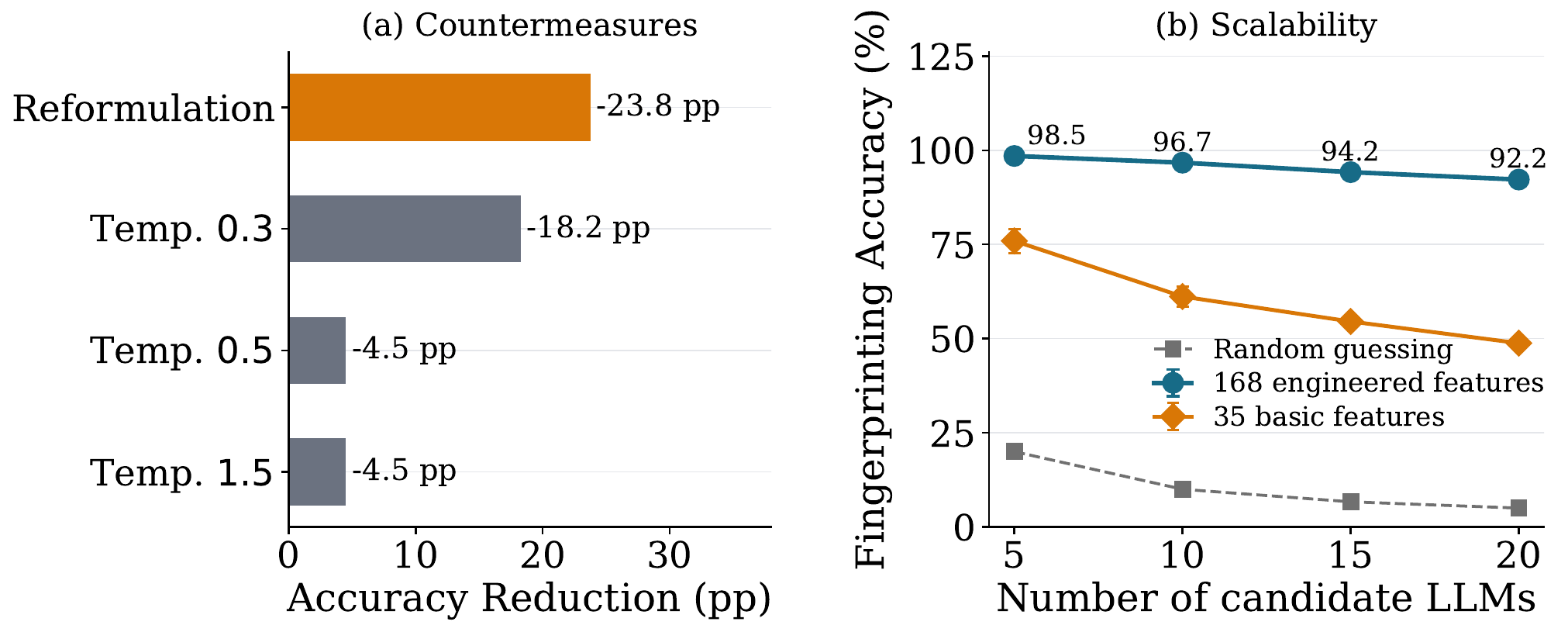}\vspace{-0.1cm}
\caption{\small (a) Accuracy reduction from prompt reformulation and
decoding-temperature randomization relative to the $93.2\%$ baseline across four prompt-grouped folds. (b) Fingerprinting accuracy for $5$--$20$ LLMs using $168$
engineered features vs. a $35$-feature baseline.} %Error bars show $95$\% confidence intervals; the dashed line denotes random guessing.}
\label{fig:fingerprinting-realistic}\vspace{-0.1in}
\end{figure}

\noindent\textbf{Network-layer defenses.}
Prior traffic-analysis and website-fingerprinting defenses use padding, token grouping, response batching, dummy packets, and traffic shaping to obscure packet-size and timing patterns. These mechanisms complement application- and model-level defenses but add bandwidth, latency, or usability overhead. Because our attacks exploit correlated signals across packet sizes, timing, bursts, rates, and response progression, effective defenses must shape the overall traffic profile rather than mask a single feature.

% Prior traffic-analysis and website-fingerprinting research proposes padding,
% token grouping, response batching, dummy packets, and traffic shaping to
% obscure packet-size and timing patterns. These mechanisms complement
% application- and model-level approaches but incur bandwidth, latency, or
% usability overhead. Since our attacks exploit multiple correlated signals,
% including packet size, timing, bursts, rates, and response progression,
% effective defenses should jointly shape the overall traffic profile rather than
% hide only a single feature.

\subsection{Future Work}
\label{sec:futureWork}
Future work should investigate defenses across all three layers. At the application layer, adversarial optimization could identify semantics-preserving prompt reformulations that minimize fingerprintability while preserving task correctness. Our prompt-reformulation results (Fig.~\ref{fig:fingerprinting-realistic}(a)) suggest that this direction is promising but underexplored. At the model-serving layer, randomized decoding, token-chunk batching, and lightweight padding merit investigation, motivated by our temperature experiments. At the network layer, adaptive traffic shaping, dummy packets, and randomized delays could jointly obscure correlated traffic signals while balancing privacy gains against latency and bandwidth overhead.

% Future work should investigate defenses across all three layers. At the
% application layer, adversarial optimization could search for
% semantics-preserving prompt reformulations that minimize traffic
% fingerprintability, analogous to adversarial robustness methods that optimize
% inputs while preserving task correctness. Our prompt-reformulation results
% (Fig.~\ref{fig:fingerprinting-realistic}(a)) suggest this direction is
% promising but requires substantially more investigation. At the model-serving
% layer, randomized decoding strategies and lightweight modifications such as
% batching or padding generated token chunks deserve further study, motivated by
% our temperature experiments (Fig.~\ref{fig:fingerprinting-realistic}(a)). At
% the network layer, adaptive traffic shaping, dummy packets, and randomized
% transmission delays could jointly obscure multiple traffic features while
% balancing privacy, latency, and bandwidth overheads.
\section{Conclusion}

We present the first systematic study and benchmark of fingerprinting across direct user--LLM and collaborative multi-agent interactions. Using encrypted traces, we show that packet sizes, directions, timing, and burst structure reveal the deployed LLM, prompt category, and collaborative task despite payload encryption. These findings establish encrypted AI traffic as a measurable metadata side channel and expose privacy risks beyond message contents. The released benchmark supports reproducible evaluation of future attacks and defenses and advances the development of AI serving systems with stronger metadata privacy.

% We presented the first systematic study and public benchmark of encrypted-traffic fingerprinting for both direct user--LLM interactions and collaborative multi-agent systems. Using a large collection of TLS-encrypted traces, we showed that observable traffic characteristics alone reveal information about deployed LLMs, prompt categories, and collaborative tasks despite transport encryption. Our results demonstrate that encrypted traffic generated by modern AI systems contains stable and measurable side-channel signals, motivating the need for privacy-preserving communication mechanisms. We hope the released benchmark enables reproducible evaluation of future fingerprinting and defense techniques and supports the development of AI serving systems with stronger metadata privacy.

% \clearpage
\bibliographystyle{IEEEtran}
\bibliography{sections/main-references}

\end{document}